\documentclass{article}
\usepackage[utf8]{inputenc}
\usepackage[margin=2.5cm]{geometry}
\usepackage{soul}
\usepackage[table,xcdraw]{xcolor}
\usepackage{graphicx}
\usepackage{multirow}
\usepackage{xspace}
\usepackage[most]{tcolorbox}
\usepackage{booktabs}
\usepackage{pifont}
\usepackage{url}
\usepackage{balance}
\usepackage{amsmath,amsfonts}
\usepackage{amssymb}
\usepackage{svg}
\usepackage{cryptocode}
\usepackage{booktabs}
\usepackage{graphicx}
\usepackage{subfig}
\usepackage{float}
\usepackage{circledsteps}

\usepackage{algorithm}
\usepackage{algpseudocode}

\usepackage[utf8]{inputenc}
\usepackage{hyperref}
\usepackage{graphicx}
\usepackage{xcolor}

\usepackage{amsmath}
\usepackage{amsthm}
\usepackage{amssymb}
\usepackage{amsfonts}
\usepackage{bbm}
\usepackage{algorithm}
\usepackage{algpseudocode}
\usepackage{mathtools}

\floatstyle{ruled}
\newfloat{algo}{htbp}{algo}
\floatname{algo}{Algorithm}

\usepackage{numbertabbing}

\usepackage{IEEEtrantools}
\allowdisplaybreaks

\usepackage[shortlabels]{enumitem}

\usepackage{tikz}
\usetikzlibrary{calc}
\usetikzlibrary{arrows}
\usetikzlibrary{arrows.meta}
\usetikzlibrary{patterns}
\usetikzlibrary{positioning}
\usetikzlibrary{decorations.pathreplacing}
\usetikzlibrary{shapes.misc}
\usetikzlibrary{spy}

\usepackage{pgfplots}
\pgfplotsset{compat=1.14}

\definecolor{myParula01Blue}{RGB}{0,114,189}
\definecolor{myParula02Orange}{RGB}{217,83,25}
\definecolor{myParula03Yellow}{RGB}{237,177,32}
\definecolor{myParula04Purple}{RGB}{126,47,142}
\definecolor{myParula05Green}{RGB}{119,172,48}
\definecolor{myParula06LightBlue}{RGB}{77,190,238}
\definecolor{myParula07Red}{RGB}{162,20,47}

\tikzset{myparula11/.style={color=myParula01Blue,solid,mark=+,mark options={solid}}}
\tikzset{myparula12/.style={color=myParula01Blue,densely dashed,mark=x,mark options={solid}}}
\tikzset{myparula13/.style={color=myParula01Blue,densely dotted,mark=o,mark options={solid}}}
\tikzset{myparula14/.style={color=myParula01Blue,dashdotted,mark=triangle,mark options={solid}}}
\tikzset{myparula15/.style={color=myParula01Blue,dashdotdotted,mark=square,mark options={solid}}}

\tikzset{myparula21/.style={color=myParula02Orange,solid,mark=+,mark options={solid}}}
\tikzset{myparula22/.style={color=myParula02Orange,densely dashed,mark=x,mark options={solid}}}
\tikzset{myparula23/.style={color=myParula02Orange,densely dotted,mark=o,mark options={solid}}}
\tikzset{myparula24/.style={color=myParula02Orange,dashdotted,mark=triangle,mark options={solid}}}
\tikzset{myparula25/.style={color=myParula02Orange,dashdotdotted,mark=square,mark options={solid}}}

\tikzset{myparula31/.style={color=myParula03Yellow,solid,mark=+,mark options={solid}}}
\tikzset{myparula32/.style={color=myParula03Yellow,densely dashed,mark=x,mark options={solid}}}
\tikzset{myparula33/.style={color=myParula03Yellow,densely dotted,mark=o,mark options={solid}}}
\tikzset{myparula34/.style={color=myParula03Yellow,dashdotted,mark=triangle,mark options={solid}}}
\tikzset{myparula35/.style={color=myParula03Yellow,dashdotdotted,mark=square,mark options={solid}}}

\tikzset{myparula41/.style={color=myParula04Purple,solid,mark=+,mark options={solid}}}
\tikzset{myparula42/.style={color=myParula04Purple,densely dashed,mark=x,mark options={solid}}}
\tikzset{myparula43/.style={color=myParula04Purple,densely dotted,mark=o,mark options={solid}}}
\tikzset{myparula44/.style={color=myParula04Purple,dashdotted,mark=triangle,mark options={solid}}}
\tikzset{myparula45/.style={color=myParula04Purple,dashdotdotted,mark=square,mark options={solid}}}

\tikzset{myparula51/.style={color=myParula05Green,solid,mark=+,mark options={solid}}}
\tikzset{myparula52/.style={color=myParula05Green,densely dashed,mark=x,mark options={solid}}}
\tikzset{myparula53/.style={color=myParula05Green,densely dotted,mark=o,mark options={solid}}}
\tikzset{myparula54/.style={color=myParula05Green,dashdotted,mark=triangle,mark options={solid}}}
\tikzset{myparula55/.style={color=myParula05Green,dashdotdotted,mark=square,mark options={solid}}}

\tikzset{myparula61/.style={color=myParula06LightBlue,solid,mark=+,mark options={solid}}}
\tikzset{myparula62/.style={color=myParula06LightBlue,densely dashed,mark=x,mark options={solid}}}
\tikzset{myparula63/.style={color=myParula06LightBlue,densely dotted,mark=o,mark options={solid}}}
\tikzset{myparula64/.style={color=myParula06LightBlue,dashdotted,mark=triangle,mark options={solid}}}
\tikzset{myparula65/.style={color=myParula06LightBlue,dashdotdotted,mark=square,mark options={solid}}}

\tikzset{myparula71/.style={color=myParula07Red,solid,mark=+,mark options={solid}}}
\tikzset{myparula72/.style={color=myParula07Red,densely dashed,mark=x,mark options={solid}}}
\tikzset{myparula73/.style={color=myParula07Red,densely dotted,mark=o,mark options={solid}}}
\tikzset{myparula74/.style={color=myParula07Red,dashdotted,mark=triangle,mark options={solid}}}
\tikzset{myparula75/.style={color=myParula07Red,dashdotdotted,mark=square,mark options={solid}}}

\theoremstyle{plain}
\newtheorem{theorem}{Theorem}
\newtheorem*{theorem*}{Theorem}

\newtheorem{lemma}{Lemma}
\newtheorem{proposition}{Proposition}

\theoremstyle{definition}
\newtheorem{definition}{Definition}

\usepackage{xspace}
\newcommand{\cf}[0]{cf.\xspace}
\newcommand{\ie}[0]{\emph{i.e.}\xspace}

\tikzset{blockchain/.style={
        x=1.25cm,
        y=1.25cm,
        node distance=0.5cm,
        block/.style = {
            minimum width=0.75cm,
            minimum height=0.75cm,
            draw,
            shade,
            top color=white,
            bottom color=black!10,
        },
        block-adv/.style = {
            block,
            bottom color=myParula07Red!50,
            draw=myParula07Red!50!black,
        },
        block-hon/.style = {
            block,
            bottom color=myParula05Green!50,
            draw=myParula05Green!50!black,
        },
        link/.style = {
            -latex,
        },
        link-adv/.style = {
            link,
        },
        link-hon/.style = {
            link,
        },
    }
}

\usepackage{pifont}% http://ctan.org/pkg/pifont
\usepackage[flushleft]{threeparttable}

\newcommand{\Thorizon}[0]{\ensuremath{T_\mathsf{hor}}}
\newcommand{\chain}[0]{\ensuremath{\mathsf{ch}}}
\newcommand{\chainconf}[0]{\ensuremath{\mathsf{Ch}}}

\newcommand{\lmdghost}[0]{\ensuremath{\operatorname{\textsc{LMD-GHOST}}}}
\newcommand{\ulmdghost}[0]{\ensuremath{\operatorname{\textsc{RLMD-GHOST}}}}
\newcommand{\ghosteph}[0]{\ensuremath{\operatorname{\textsc{GHOST-Eph}}}}
\newcommand{\ghost}[0]{\ensuremath{\operatorname{\textsc{GHOST}}}}

\newcommand{\filteredghost}[0]{\ensuremath{\mathcal{G}_f}}
\newcommand{\Hinf}[0]{\ensuremath{\widetilde{H}}}

\long\def\blockcomment#1\endblockcomment{}
\algdef{SE}[DOWHILE]{Do}{doWhile}{\algorithmicdo}[1]{\algorithmicwhile\ #1}%

\algdef{SE}[UPON]{Upon}{EndUpon}[1]{\textbf{upon}\ $#1$}{\textbf{end upon}}%
\algdef{SE}[AT]{At}{EndAt}[1]{\textbf{at}\ $#1$}{\textbf{end at}}%

\newcommand{\GAT}[0]{\ensuremath{\mathsf{GAT}}}
\newcommand{\GST}[0]{\ensuremath{\mathsf{GST}}}
\newcommand{\TPA}[0]{\ensuremath{\mathsf{tpa}}}
\newcommand{\varTPA}[1]{\ensuremath{\mathsf{#1\text{-}tpa}}}

\newcommand{\piTPA}[0]{\ensuremath{\mathsf{\pi\text{-}tpa}}}
\newcommand{\V}[0]{\ensuremath{\mathcal{V}}}

\newcommand{\FC}[0]{\ensuremath{\mathsf{FC}}}
\newcommand{\negl}[0]{\ensuremath{\operatorname{negl}}}

\newcommand{\Tconf}[0]{\ensuremath{T_\mathsf{conf}}}

\newcommand{\B}[0]{\ensuremath{\mathcal{B}}}

\newcommand{\node}[0]{\ensuremath{\mathcal{P}}}

\newcommand{\Goldfish}[0]{\textsf{Goldfish}\xspace}
\newcommand{\LMDGHOST}[0]{\textsf{LMD-GHOST}\xspace}
\newcommand{\ULMDGHOST}[0]{\textsf{RLMD-GHOST}\xspace}

\newcommand{\FIL}[0]{\textsf{FIL}\xspace}

\newcommand{\LOGbft}[2]{%
    \ifthenelse{\equal{#1}{}}{%
        \ensuremath{\mathsf{LOG}_{\mathrm{bft}}^{#2}}%
    }{%
        \ensuremath{\mathsf{LOG}_{\mathrm{bft},#1}^{#2}}%
    }%
}

\newcommand{\ld}[1]{%
    \ifthenelse{\equal{#1}{}}{%
        \ensuremath{\mathrm{L}^{(c)}}%
    }{%
        \ensuremath{\mathrm{L}^{(#1)}}%
    }%
}

\newcommand{\bprop}[1]{%
    \ifthenelse{\equal{#1}{}}{%
        \ensuremath{\Hat{b}}%
    }{%
        \ensuremath{\Hat{b}_{#1}}%
    }%
}

\title{Recent Latest Message Driven GHOST: Balancing Dynamic Availability With Asynchrony Resilience}
\author{Francesco D'Amato\\
  Ethereum Foundation\\
  \url{francesco.damato@ethereum.org}
  \and Luca Zanolini\\
   Ethereum Foundation\\
  \url{luca.zanolini@ethereum.org}
}

\date{}

\begin{document}
\maketitle
%\tableofcontents

\begin{abstract}\noindent
Dynamic participation has recently become a crucial requirement for devising permissionless consensus protocols. This notion, originally formalized by Pass and Shi (ASIACRYPT 2017) through their ``sleepy model", captures the essence of a system's ability to handle participants joining or leaving during a protocol execution. A dynamically available consensus protocol preserves safety and liveness while allowing dynamic participation. Blockchain protocols, such as Bitcoin's consensus protocol, have implicitly adopted this concept.

In the context of Ethereum's consensus protocol, Gasper, Neu, Tas, and Tse (S\&P 2021) presented an attack against LMD-GHOST -- the component of Gasper designed to ensure dynamic availability. Consequently, LMD-GHOST results unable to fulfill its intended function of providing dynamic availability for the protocol. Despite attempts to mitigate this issue, the modified protocol still does not achieve dynamic availability, highlighting the need for more secure dynamically available protocols.

In this work, we present RLMD-GHOST, a synchronous consensus protocol that not only ensures dynamic availability but also maintains safety during bounded periods of asynchrony. This protocol is particularly appealing for practical systems where strict synchrony assumptions may not always hold, contrary to general assumptions in standard synchronous protocols. 

Additionally, we present the ``generalized sleepy model", within which our results are proven. Building upon the original sleepy model proposed by Pass and Shi, our model extends it with more generalized and stronger constraints on the corruption and sleepiness power of the adversary. This approach allows us to explore a wide range of dynamic participation regimes, spanning from complete dynamic participation to no dynamic participation, i.e., with every participant online. Consequently, this model provides a foundation for analyzing dynamically available protocols.
\end{abstract}

\section{Introduction}
\label{sec:introduction}

\subsection{Balancing dynamic availability and asynchrony resilience}

Tolerating \emph{dynamic participation} has emerged as a desirable feature of consensus protocols operating in the permissionless setting of blockchains. A \emph{dynamically available} protocol preserves safety and liveness during events involving portions of the participants going offline. This concept, which was implicitly taken into account by the Bitcoin consensus protocol~\cite{nakamoto2008bitcoin}, was later formalized by Pass and Shi~\cite{sleepy} through the \emph{sleepy model} of consensus. In particular, Pass and Shi model a system where participants can be either online or offline, with their online status adversarially controlled throughout the execution, subject to the honest and online participants always outnumbering the adversarial ones.  

A notable limitation of dynamically available consensus protocols is their inability to tolerate network partitions~\cite{DBLP:journals/sigact/GilbertL02, sleepy, cap2}. No consensus protocol can simultaneously satisfy liveness (under dynamic participation) and safety (under temporary network partitions, or temporary asynchrony). In other words, it is impossible for a consensus protocol (for state-machine replication) to yield a single output chain, while simultaneously offering dynamic availability and ensuring transaction finality, even during asynchronous periods or network partitions. As a result, dynamically available protocols are generally assumed to be synchronous~\cite{sleepy, DBLP:conf/ccs/Momose022, DBLP:journals/iacr/MalkhiMR22}. 

Neu, Tas, and Tse~\cite{DBLP:conf/sp/NeuTT21}, while formalizing the security requirements of Ethereum's consensus protocol, Gasper~\cite{gasper}, demonstrate that the original version of \LMDGHOST, Gasper's dynamically available component, is not secure even in a context of full participation, \ie, with all the participants in the protocol, or \emph{validators}, being online. This finding is supported by the presentation of a \emph{balancing} attack~\cite{DBLP:conf/sp/NeuTT21, DBLP:conf/fc/Schwarz-Schilling22}.%in which the adversary carefully times the release of messages in order to get honest validators to vote on two different chains. This way, it can maintain the block tree in a \emph{balanced}, unresolved state, where neither of the two chain ever definitely wins out. The proposer boost technique~\cite{boost} was later introduced as a mitigation, in order to help \emph{coordinate the voters in slots with an honest proposer}, thus preventing the adversary from splitting honest votes. Still, the resulting protocol falls short of achieving dynamic availability, and moreover it is quite prone to reorgs.

D'Amato, Neu, Tas, and Tse~\cite{goldfish} have proposed \Goldfish as a solution to address the challenges posed by \LMDGHOST. \Goldfish is synchronous consensus protocol that offers safety and liveness even when there is variable participation, making it dynamically available. However, the protocol's lack of \emph{resilience to temporary asynchrony} makes it impractical to replace \LMDGHOST in Ethereum: even a very short period of asynchrony can result in a catastrophic failure, jeopardizing the safety of any previously confirmed block.

Our work acknowledges the limitations of both \LMDGHOST and \Goldfish, and proposes a new family of synchronous consensus protocols, namely Recent Latest Message Driven GHOST (\ULMDGHOST), generalizing (variants of) \LMDGHOST and \Goldfish. To analyze its properties, we introduce the \emph{generalized sleepy model}, which can capture a broad spectrum of dynamic participation regimes falling between complete dynamic participation and no dynamic participation, and within which we formally define the concept of \emph{asynchrony resilience}. Through the \ULMDGHOST family, we explore the trade-off space between resilience to temporary asynchrony and dynamic availability.

\subsection{Technical outline}
\label{sec:outline}

\subsubsection{\LMDGHOST and its security}
The consensus protocol of Ethereum, Gasper~\cite{gasper}, is defined by two \emph{components}~\cite{DBLP:conf/sp/NeuTT21}: \LMDGHOST, a synchronous consensus protocol, and Casper~\cite{casper}, a partially synchronous protocol (or \emph{gadget}) that finalizes blocks on the chain output by \LMDGHOST and that keeps such finalized blocks safe during period of network asynchrony. 

 \LMDGHOST is named after its key component, the Latest Message Driven Greediest Heaviest Observed Sub-Tree rule ($\lmdghost$) \emph{fork-choice function} (Section~\ref{sec:lmd-ghost}),  introduced by Zamfir~\cite{zamfir}. This function takes as input the sequence of blocks and the votes from each validator $v_i$, \ie, a \emph{local view} of $v_i$, and outputs a \emph{canonical chain}. To do so, it starts from the genesis block $B_{\text{genesis}}$ and walks down the sequence of blocks: at each block $B$, it chooses as the next block the child of $B$ with the \emph{heaviest subtree}, \ie, with most (stake weighted) \emph{latest} votes on its descendants. 
 
 \LMDGHOST can be thought of as a \emph{propose-vote} protocol, proceeding in \emph{slots}, each with an associated validator, a \emph{proposer}, tasked with creating and proposing a new block. Votes are then cast by a randomly selected subset of the validator set, a \emph{committee}. In other words, the protocol implements \emph{subsampling} of validators. Both the proposer and voters utilize the output of the $\lmdghost$ fork-choice function to decide what block to extend, and what block to vote for, respectively.

\paragraph{Balancing attack}
Neu, Tas, and Tse~\cite{DBLP:conf/sp/NeuTT21} show that this protocol is vulnerable to a \emph{balancing attack}~\cite{DBLP:conf/sp/NeuTT21, DBLP:conf/fc/Schwarz-Schilling22}, exploiting the fact that validators can have different local views, leading to conflicting votes. An adversarial proposer of slot $t$ can equivocate and produce two conflicting blocks, which it reveals to two equal-sized subsets of the honest validators in the committee of slot $t$, so that their votes are split between the two blocks. In slot $t+1$, it then releases withheld votes from the adversarial validators in the committees of slot $t$ to split validators of slot $t+1$ into two equal-sized subsets, one which sees one chain as leading and votes for it, and one which sees the other chain as leading and votes for it. This can be repeated indefinitely, maintaining the split between the two chains and preventing either liveness or safety of \LMDGHOST, as any decision made while the attack is ongoing would not be safe. The proposer boost technique~\cite{boost} was later introduced as a mitigation to help \emph{coordinate the voters in slots with an honest proposer}. It requires honest voters to temporarily grant extra weight to the current proposal. 

\paragraph{Ex-ante reorgs}
Still, the resulting protocol remains prone to \emph{ex-ante reorgs}~\cite{DBLP:conf/fc/Schwarz-Schilling22}, and has not been proven secure, even in a regime of full participation. Consider an adversary controlling a fraction $\beta$ of the validators, among which the proposer for slot $t+1$, which \emph{privately} creates a block $B'$ on top of block $B$, \ie, the block for slot $t$. Moreover, the adversary controls roughly a fraction $\beta$ of the validators sampled to vote at slot $t+1$, and they privately vote for $B'$. Honest validators in the committee of slot $t+1$ do not see any block and thus vote for $B$. In the next slot, \ie, slot $t+2$, an honest proposer publishes block $B''$ building on $B$, which is the current tip of the canonical chain in their local view, and all honest validators in the committee of slot $t+2$ vote for it, so $B''$ accrues roughly $1-\beta$ of one committee's weight, say $(1-\beta)W_c$. Afterwards, the adversary publishes block $B'$ and its votes from slot $t+1$ \emph{and} $t+2$ for it. At this point, $B$ has weight roughly $2\beta W_c$. If $2\beta > 1-\beta$, \ie, if $\beta > \frac{1}{3}$, $B'$ becomes canonical, reorging $B''$. More generally, an adversary controlling $k$ slots in a row (which happens with probability $\beta^k$) can withhold votes from those slots, and release them after one honest slot, resulting in $(k+1)\beta W_c$ weight for a withheld adversarial block. If $\beta > \frac{1}{k+2}$, this result in a reorg of the honestly proposed block, so even relatively weak adversaries can perform ex-ante reorgs, albeit with low probability.

Note that subsampling is crucial in the success of the attack, because the adversary is able to accumulate the votes of the validators it controls across the $k$ committees. Without subsampling, the adversarial votes from the last slot would simply replace the ones from the previous slots, due to the latest message rule. In other words, \emph{without subsampling there is no way for the adversary to overcome the weight of all honest validators voting together in one slot}, as they are a majority over the entire validator set. We will later make use of this fact in our protocol.

%\paragraph{Dynamic availability failure due to stale votes}
%Finally, while the security of \LMDGHOST is not completely clear under full participation, it is certainly insecure under dynamic participation. Consider a validator set partitioned in three sets, $V_1$, $V_2$, and $V_3$, with $|V_1| = |V_2| + 1$ and $V_3$ much smaller. Validators in $V_1$ and $V_2$ are honest, while those in $V_3$ are adversarial. Let~$t-1$ and~$t$ be two slots with adversarial proposers. In slot~$t$, the adversary publishes conflicting blocks $B$ and $B'$, one as a proposal for slot $t-1$ and the other for slot~$t$, in such a way that validators in $V_1$ only receive $B$ before voting, and thus vote for $B$, and validators in $V_2$ only see $B'$, and vote for it. $B$ then becomes canonical, since $|V_1| > |V_2|$. Now, validators in $V_2$ go offline. This is allowed by the sleepy model, since validators in $V_1$ are online and still outnumber the adversarial validators. The adversary can now wait arbitrarily long and then release votes for $B'$, resulting in a reorg from $B$ to $B'$, since $|V_2| + |V_3| > |V_1|$ and the latest votes of $V_2$ are those for $B'$, from slot $t$, which are still counted. If no decision has been made in the meantime, the protocol has failed to be live under dynamic participation. If a decision has been made, it must be for a descendant of $B$, and we have a safety failure. We describe this attack and its consequences more carefully in Theorem~\ref{thm:lmd-not-da}.

\subsubsection{\Goldfish}

To cope with the problems of \LMDGHOST, D'Amato, Neu, Tas, and Tse~\cite{goldfish} devise \Goldfish (Section~\ref{sec:goldfish}), a synchronous consensus protocol that enjoys safety and liveness \emph{under fully variable participation}, and thus that is dynamically available. Moreover, \Goldfish is \emph{reorg resilient}: blocks proposed by honest validators are guaranteed inclusion in the chain. \Goldfish is based on two techniques: \emph{view-merge}~\cite{ethresearch-view-merge-2, highway}\footnote{View-merge was first introduced in the Highway protocol~\cite{highway}, under the name of vote buffering, which is also the name used in \Goldfish~\cite{goldfish}.} and \emph{vote expiry}. It inherits the propose-vote structure of \LMDGHOST, and it adds a third round dedicated to view-merge. This new structure can be generalized by a family of protocols which we call \emph{propose-vote-merge} protocols (Section~\ref{sec:pvm}).

View-merge is an improvement over the proposer boost technique, and serves the same purpose, \ie, coordinating honest voters in slots with an honest proposer. In particular, it lets an honest proposer synchronize the local views of honest voters with its own, and \emph{ensures} that they cast votes for the proposed block. With vote expiry, only votes from the latest slot influence the fork-choice. The adversary has then no way to accumulate weight \emph{across committees}, and thus \emph{no way to overpower the coordinated voting of the honest voters in a committee}. Therefore, vote expiry solves two problems at once: the ex-ante reorgs caused by subsampling, and the lack of dynamic availability due the adversary exploiting votes of offline validators, as these are now simply expired. 

As we now informally explain, view-merge and vote expiry together make for a reorg resilience protocol, under the assumption that in every committee a majority of all \emph{online} validators are honest\footnote{The sleepy model assumes that at any given time a majority of online validators are honest, which is a slightly different assumption that assuming such condition for a committee. Still, the latter can be recovered by strengthening the former, by requiring a $\frac{1}{2} + \epsilon$ majority in the whole validator set, and by having large enough committees so that an honest majority in a committee holds except with negligible probability.}. View-merge ensures that an honest proposal $B$ from slot $t$ receives \emph{all} votes from \emph{honest and online} validators in the committee of slot $t$, which by assumption outnumber the adversarial votes in it. By vote expiry, the votes from the committee of slot $t$ are the only ones which count in slot $t+1$, \ie, voters in slot $t+1$ use them as input to their fork-choice function when determining what to vote for. Since a majority of the votes which are considered are for $B$, all votes from honest and online validators in slot $t+1$ are for $B$ as well. By induction, this is true for all future slots achieving reorg resilience of honest proposals. As we explore in Section~\ref{sec:pvm}, reorg resilience then immediately implies that the $\kappa$-deep confirmation rule is both safe and live, implying dynamic availability for \Goldfish. 

Nonetheless, \Goldfish is not considered practically viable to replace \LMDGHOST in Ethereum, due to its brittleness to temporary asynchrony: even a single slot of asynchrony can lead to a catastrophic failure, jeopardizing the safety of \emph{any} previously confirmed block. In other words, \Goldfish is not \emph{asynchrony resilient}. This is simply due to the strict vote expiry, which can cause the weight supporting a block to drop to zero at any time, if no \emph{new} votes supporting it are received within one slot. 

\subsubsection{\ULMDGHOST}

Strict vote expiry does not seem compatible with any reasonable notion of resilience to asynchrony. On the other end, not expiring votes at all is not compatible with dynamic availability, as we have observed in \LMDGHOST, where votes of \emph{honest but offline} validators can be exploited by the adversary to conclude long reorgs. A natural approach, and the one we take in this work, is then to \emph{relax} vote expiry. In particular, we introduce Recent Latest Message Driven GHOST (\ULMDGHOST) (Section~\ref{sec:ULMD}), \emph{family} of propose-vote-merge consensus protocols parameterized by the \emph{vote expiry period $\eta$}, \ie, only votes from the most recent $\eta$ slots are utilized in the protocol with parameter $\eta$. \ULMDGHOST extends and generalizes both \LMDGHOST and \Goldfish. As \LMDGHOST, \ULMDGHOST implements the latest message rule (LMD). As \Goldfish, it implements view-merge and vote expiry. Relaxing vote expiry forces us to do away with subsampling, since we want to avoid reintroducing the possibility of ex-ante reorgs through accumulating votes from multiple committees, as this would break the reorg resilience property upon which the security of both \Goldfish and \ULMDGHOST rests (and which is itself highly desirable). Therefore, every validator votes in every slot of \ULMDGHOST. For $\eta = 1$, \ULMDGHOST reduces to a variant of \Goldfish without subsampling of validators. For $\eta = \infty$, \ULMDGHOST reduces to \LMDGHOST. To be precise, we do not refer here to \LMDGHOST as currently implemented in Ethereum but to a variant, described in this work, without subsampling \emph{and implementing the view-merge technique instead of proposer boost} (\ie, also a propose-vote-merge protocol).

Intuitively, considering votes from a longer period ($\eta > 1$) results in a protocol that is more tolerant of asynchrony, since a correspondingly longer period of asynchrony is needed for all votes to expire. On the other end, relaxing vote expiry should weaken dynamic availability because, as in \LMDGHOST, the fork-choice can now be affected by the (still unexpired) votes of offline validators. In the reorg resilience argument for \Goldfish, after an honest slot $t$ with proposal $B$, honest voters at slot $t+1$ still vote for $B$, because the honest votes from slot $t$ outnumber adversarial votes, \emph{and there are no further votes to consider}. With a longer expiry period $\eta > 1$, this is not the case anymore, because at slot $t+1$ all votes from slots $[t+1-\eta, t]$ have to be considered.

\subsubsection{Generalized sleepy model}

In order to describe the security of \ULMDGHOST, we introduce an extension of the sleepy model, which we refer to as the \emph{generalized sleepy model} (Section~\ref{sec:model}), that allows us to capture a weaker notion of dynamic availability and to precisely define a notion of \emph{asynchrony resilience}. In defining this model, we again take inspiration from the reorg resilience argument for \Goldfish. In particular, computing the fork-choice at slot $t$ now requires considering all votes from slots $[t-\eta, t-1]$. If there has been an honest proposal in slot $t-1$, the only votes which are guaranteed to vote in support of it are those from honest validators which were online in slot $t-1$, which we define as $H_{t-1}$. All other votes are either adversarial or from offline validators, and potentially dangerous in both cases. Letting $H_{s,t}$ be the honest validators online in slots $[s,t]$, and $A_t$ be the validators corrupted up to slot $t$, we require that $|H_{t-1}| > |A_t \cup (H_{t-\eta, t-2}\setminus H_{t-1})|$. Observe that when $\eta = 1$, we recover a standard majority assumption, \ie, $|H_{t-1}| > |A_t|$, much like the one found in the sleepy model. The more general condition is precisely the key assumption of our \emph{generalized sleepy model}, which allows us to describe a whole spectrum of dynamic participation regimes, by varying the parameter. If the assumption holds for a certain $\eta$, then the corresponding \ULMDGHOST protocol is reorg resilient, and consequently secure. Finally, we extend the generalized sleepy model further, to allow for bounded periods of asynchrony, and define an associated notion of asynchrony resilience, which is satisfied by \ULMDGHOST. Limitations of \ULMDGHOST are analyzed in Appendix~\ref{appendix:limitations}.

Related works to this paper are discussed in Section~\ref{sec:related-works}, while conclusion and future work are presented in Section~\ref{sec:conclusion}. Finally, in Appendix~\ref{appendix-fast-conf} we extend \ULMDGHOST achieving a faster confirmation time for proposals, \emph{optimistically}. This is aligned with the way in which \Goldfish also achieves fast confirmation, with the difference that we do not require an increase in the length of the slots, due to using a slightly different proposer selection mechanism.

\section{Model and Preliminary Notions}
\label{sec:model}

\paragraph*{Validators}
We consider a system of $n$ \emph{validators} $v_1, \dots, v_n$ that communicate with each other through exchanging messages. Every validator is identified by a unique cryptographic identity and the public keys are common knowledge. Validators are assigned a protocol to follow, consisting of a collection of programs with instructions for all validators.

\paragraph*{Failures}
A validator that follows its protocol during an execution is called \emph{honest}. On the other hand, a faulty validator may crash or even deviate arbitrarily from its specification, e.g., when corrupted by an adversary. We consider Byzantine faults here and assume the existence of a probabilistic poly-time adversary $\mathcal{A}$ that can corrupt validators over the course of the entire protocol execution. Corrupted validators stay corrupted for the remaining duration of the protocol execution, and are thereafter called \emph{adversarial}. The adversary $\mathcal{A}$ knows the the internal state of adversarial validators. The adversary is \emph{adaptive}: it chooses the corruption schedule dynamically, during the protocol execution.

\paragraph*{Links}

We assume that a best-effort gossip primitive that will reach all validators is available. Moreover, we assume that messages from honest validator to honest validator are eventually received and cannot be forged. This includes messages sent by Byzantine validators, once they have been received by some honest validator $v_i$ and gossiped by $v_i$.

\paragraph*{Time} 
{Time is divided into discrete \emph{rounds}. We consider a synchronous model in which validators have synchronized clocks and message delays are bounded by $\Delta$ rounds. }Moreover, we define the notion of \emph{slot} as a collection of $k$ rounds, for a constant $k$. We are interested in the case $k=3\Delta$, so our presentation will assume this length for slots, unless otherwise specified.

\paragraph*{Sleepiness}
The adversary $\mathcal{A}$ can decide for each round which honest validator is \emph{awake} or \emph{asleep} at that round. Asleep validators do not execute the protocol and messages for that round are queued and delivered in the first round in which the validator is awake again. Honest validators that become awake at round $r$, before starting to participate in the protocol, must first execute (and terminate) a \emph{joining protocol} (see Section~\ref{sec:joining-protocol}), after which they become \emph{active}~\cite{goldfish}.
All adversarial validators are always awake, and are not prescribed to follow any protocol. Therefore, we always use active, awake, and asleep to refer to honest validators. As for corruptions, the adversary is adaptive also for sleepiness, \ie, the sleepiness schedule is also chosen dynamically by the adversary. Note that awake and active validators coincide in the sleepy model~\cite{sleepy}.

\paragraph*{Proposer election mechanism}
In each slot $t$, a validator~$v_p$ is selected to be a \emph{proposer} for~$t$, \ie, to extend the chain with a new block. Observe that, when we want to highlight the fact that $v_p$ is a proposer for a specific slot~$t$, we use the notation~$v_{p}^t$. Otherwise, when it is clear from the context, we just drop the slot $t$, to make the notation simpler. As the specification of a proposal mechanism is not within the goals of this work, we assume the existence of a proposer selection mechanism satisfying the requirements of a Single Secret Leader Election (SSLE) scheme~\cite{ssle}, \ie, \emph{uniqueness, unpredictability, and fairness}: $v_p$ is unique, the identity of~$v_p$ is only known to other validators once~$v_p$ reveals itself, and any validator has probability~$\frac{1}{n}$ of being elected to be a proposer at any slot. Such a mechanism has been researched for usage in the Ethereum consensus protocol~\cite{ethresearch-whisk}. 

\paragraph*{View} Due to adversarial validators and message delays, validators may have different sets of received messages. A \emph{view} (at a given round $r$), denoted by $\V$, is a subset of all the messages that a validator has received until $r$. Observe that the notion of view is \emph{local} for the validators. For this reason, when we want to focus the attention on a specific view of a validator $v_i$, we denote with $\V_i$ the view of $v_i$ (at a round $r$). There are validity conditions on messages, and we say that a view is valid if all messages within it are \emph{verifiably valid within the view itself}, \ie, all messages they reference are also contained in the view and themselves valid. In particular, a block contains a reference to its parent block, and verifying its validity requires also being able to verify the parent's validity (and recursively that of the entire chain). We do not discuss questions of availability and validity further, and just leave it implicit that we only ever consider valid messages and views.

\paragraph*{Blocks and chains} For two chains $\chain_1$ and $\chain_2$, we say $\chain_1 \preceq \chain_2$ if $\chain_1$ is a prefix of $\chain_2$. If block $B$ is the tip of chain $\chain$, we say that it is the \emph{head of $\chain$}, and we identify the whole chain with $B$. Accordingly, if $\chain' \preceq \chain$ and $A$ is the head of $\chain'$, we also say $\chain' \preceq B$ and $A \preceq B$.

\paragraph*{Fork-choice functions} A fork-choice function is a deterministic function $\FC$, which takes as input a view $\V$ and a slot $t$ and outputs a block $B$, satisfying the following \emph{consistency property}: if $B$ is a block extending $\FC(\V, t)$, then $\FC(\V \cup \{B\}, t) = B$. We refer to the output of $\FC$ as the \emph{head of the canonical chain in $\V$}, and to the chain whose head is $B$ as the \emph{canonical chain in $\V$}. Each validator keeps track of its canonical chain, which it updates using $\FC$, based on its local view. We refer to the canonical chain of validator $v_i$ at round~$r$ as $\chain_i^r$. In this work we are mainly interested in a particular class of fork-choice functions based on $\ghost$~\cite{ghost}, which we denote with $\filteredghost$ and introduce in Section~\ref{sec:ghost-based-fc}. 

\paragraph*{Terminology}
We often use the terms \emph{honest proposal}, \emph{honest slot}, and \emph{honest view} to refer to a block proposal made by an honest validator, a slot with an honest proposer, and a view of an honest validator, respectively. We also use the term \emph{pivot slot} to refer to a slot in which the proposer is active at proposal time, \ie, to a slot in which an honest proposal is made, and we say that such a slot has an \emph{active proposer}. Finally, we say \emph{honest voters of slot $t$} to refer to the active validators at the voting round $3\Delta t + \Delta$ of slot~$t$.

\subsection{Generalized sleepy model}

We now specify how the adversary is constrained in using its corruption and sleepiness power. We do so by formulating first a one-parameter family of adversarial restrictions, which generalizes the usual sleepy model~\cite{sleepy, goldfish}, and then a two-parameters family, which generalizes it further by introducing, and accounting for, \emph{bounded} periods of asynchrony.

\subsubsection{$\tau$-sleepy model}

We denote with $h_r$ the number of honest validators that are active at round $r$, with $h_0 > 0$ a lower bound on $h_r$, and with $f_r$ the number of adversarial validators at round $r$. In the sleepy model~\cite{sleepy}, the adversary is constrained in its choice of sleepiness and corruption schedules by the requirement that awake validators outnumber adversarial validators in every round {by a constant factor $c > 1$}. As awake and active validators coincide in this model, the requirement is $h_r > cf_r$.

D'Amato, Neu, Tas, and Tse~\cite{goldfish} introduce the notion of an active validator\footnote{There, validators which have completed the joining protocol are simply called awake, and validators which are executing the joining protocol are called dreamy.} and assume a modified condition, \ie, $h_{r - 3\Delta} > f_{r}$. In this condition, that is tailored to their protocol, \Goldfish, $h_{r-3\Delta}$ is considered instead of $h_r$ because, if~$r$ is a voting round in \Goldfish, validators corrupted after round~$r$ can still retroactively broadcast votes for that round, and these votes are relevant until $3\Delta$  rounds later (but no longer, due to vote expiry). In practice, all that is required is that $h_{3\Delta (t-1) + \Delta} > f_{3\Delta t + \Delta}$ for any \emph{slot} $t$, \ie, the condition only needs to hold for \emph{voting rounds}.

In this work, we follow this distinction between awake and active validators, and we use $H_t$ and $A_t$, for a slot $t$, to refer to the set of active and adversarial validators at round $3\Delta t + \Delta$, respectively \footnote{Recall that we focus on protocols of slot length $3\Delta$ rounds. For the protocols introduced in Section~\ref{sec:pvm}, $3\Delta t + \Delta$ is a voting round, as in \Goldfish.}. Moreover, we define $H_{s, t}$ as the set of validators that are active \emph{at some point} in slots $[s,t]$, \ie, $H_{s,t} = \bigcup_{i=s}^t H_i$ (if $i < 0$ then $H_i \coloneqq \emptyset$). We then require that, for some fixed parameter $1\leq \tau \leq \infty$, the following condition, which we refer to as  \emph{$\tau$-sleepiness at slot $t$}, holds for any slot $t$: 
\begin{equation}
\label{eq:sleepy-req}
  |H_{t-1}| > |A_{t} \cup (H_{t-\tau, t-2}\setminus H_{t-1})|
\end{equation}
{Equation~\ref{eq:sleepy-req} requires that the number of active validators at slot~$t-1$, \ie, $|H_{t-1}|$, must exceed the number of adversarial validators, \ie, $|A_{t}|$, together with all the other validators that have been active at some point in $[t-\tau,t-2]$, and that are not active at slot $t-1$, \ie, $|H_{t-\tau, t-2}\setminus H_{t-1}|$.} {Intuitively, this condition is tailored to a protocol implementing vote expiry (with period $\eta = \tau$) because the votes which are considered at slot $t$ are those from slots $[t-\tau, t-1]$. Out of these, the only honest votes which we can rely on are those from $H_{t-1}$, whereas unexpired votes from honest validators which were not active in slot $t-1$ might help the adversary.}

We refer to the sleepy model in which the adversary is constrained by $\tau$-sleepiness as the \emph{$\tau$-sleepy model}. Note that, for $\tau = 1$, this reduces to the sleepy model from \Goldfish, as this condition reduces to the majority condition $h_{r - 3\Delta} > f_{r}$ of \Goldfish for voting rounds $r = 3\Delta t + \Delta$, because $H_{t-1, t-2} = \emptyset$. We therefore also refer to the $1$-sleepy model simply as sleepy model.

\begin{definition}[$\tau$-compliant execution]
An execution is \emph{$\tau$-compliant} if it satisfies $\tau$-sleepiness. We refer to the set of such protocol executions as $E_{\tau}$. In other words, the $\tau$-sleepy model restricts the allowable set of executions to $\tau$-compliant executions, \ie, to $E_{\tau}$, constraining the adversarial sleepiness and corruption power accordingly. We refer to $1$-compliant executions simply as compliant executions. 
\end{definition}

\paragraph*{Hierarchy of $\tau$-sleepy models}
As $\tau$ increases, so do the restrictions that $\tau$-sleepy models put on the adversary, \ie, the $\tau_1$-sleepy model makes stronger assumptions than the $\tau_2$-sleepy model for $\tau_1 > \tau_2$. Another way to say this is that $\tau_1 > \tau_2$ implies $E_{\tau_1} \subset E_{\tau_2}$. This is immediate from $\tau$-sleepiness, \ie, Equation~\ref{eq:sleepy-req}. The only term that depends on $\tau$ is $|H_{t-\tau, t-2}\setminus H_{t-1}|$, which is monotonically increasing in $\tau$. Therefore, $\tau_1$-sleepiness implies $\tau_2$-sleepiness, so a $\tau_1$-compliant execution is also a $\tau_2$-compliant execution. In other words, increasing $\tau$ makes it harder for $\tau$-sleepiness to be satisfied, \ie, it constrains the adversarial corruption and sleepiness power more. 

As we mentioned, $\tau = 1$ corresponds to  sleepy model from \Goldfish, which constrains the adversary in the minimum way that can allow for a secure protocol under dynamic participation. For $\tau = \infty$, $\tau$-sleepiness requires that $|H_{t-1}| > |A_{t} \cup (H_{0, t-2}\setminus H_{t-1})|$, \ie, all honest validators which are not active at round $3\Delta (t-1) + \Delta$, and which have voted at least once in the past, are counted together with the adversarial ones. If all validators have voted at least once in slots $[0,s-1]$, this requires that $|H_t| > \frac{n}{2}$ for all slots~$t > s$, \ie, dynamic participation is allowed only in an extremely narrow sense.

\subsubsection{$(\tau, \pi)$-sleepy model}

We generalize the $\tau$-sleepy model further, by introducing the notion of a \emph{temporary period of asynchrony of less than $\pi$ slots}, abbreviated by $\piTPA$. In particular, we define a temporary period of asynchrony as it follows.

\begin{definition}[Temporary period of asynchrony] 
We say that an interval $(t_1, t_2)$ of consecutive slots is a \emph{temporary period of asynchrony}, abbreviated by $\TPA$, if synchrony does not hold in $(t_1, t_2)$. If $t_2 - t_1 \leq \pi$, we also refer to it as $\piTPA$.
\end{definition}

We consider a system where synchrony holds \emph{except for one such $\piTPA$}, for some $\pi \in \mathbb{N} \cup \{\infty\}$. We refer to this network model as \emph{synchronous network with a temporary period of asynchrony $\TPA$}. Since a $1$-$\TPA$ is empty, this is a generalization of the usual synchronous network model. We also specify a suitable notion of compliance for executions in this network model, which defines the \emph{$(\tau, \pi)$-sleepy model}, generalizing the $\tau$-sleepy model. 

\begin{definition}[($\tau, \pi$)-compliant execution]
\label{def:tau-pi-compliance}
For $\tau > \pi$, or $\tau = \pi = \infty$, an execution in the synchronous network model with a $\TPA$ is $(\tau, \pi)$-compliant if the $\TPA$ is in particular a $\piTPA \ (t_1, t_2)$ and the following conditions hold:
\begin{itemize}
    \item $\tau$-sleepiness at slot $t$ holds for $t \not \in (t_1, t_2]$ 
    \item $|H_{t_1} \setminus A_{t}| > |A_{t} \cup (H_{t-\tau, t-1}\setminus H_{t_1})|$ for $t \in (t_1, t_2+1]$
    \item $H_{t_1}$ are awake at round $3\Delta t_1 + 2\Delta$
\end{itemize}
We say that a $(\tau, \pi)$-compliant execution satisfies $(\tau, \pi)$-sleepiness, and call the set of $(\tau, \pi)$-compliant executions $E_{\tau, \pi}$. The $(\tau, \pi)$-sleepy model restricts the allowable set of executions to $E_{\tau, \pi}$. 
\end{definition}

During the $\piTPA$, the network is asynchronous and \emph{all} honest validators can be asleep. On the other hand, there are more restrictions on the adversary corruption schedule, \ie, the adversary cannot corrupt too many validators in $H_{t_1}$, because we rely on $H_{t_1}$ to preserve the canonical chain throughout this period. This is also why we have the third condition, as it guarantees that validators $H_{t_1}$ are able to observe the votes cast at round $3\Delta t_1 + \Delta$, which inform any votes they might cast during the period of asynchrony.
Moreover, not too many honest validators can be woken up during this period, because waking up during asynchrony allows the adversary to manipulate their votes. Note that, $\forall t > t_2$, $\tau$-sleepiness holds at slot $t$, and the network is synchronous. Unless otherwise specified, we will mainly consider the $\tau$-sleepy model. The $(\tau, \pi)$-sleepy model will be used when interested in analyzing the behaviour of a protocol under (bounded) asynchrony. In particular, we use it to define what it means for a synchronous protocol to be \emph{resilient to (temporary) asynchrony} (Definition~\ref{def:asynchrony-resilience}). 

\paragraph*{Hierarchy of $(\tau, \pi)$-sleepy models}
Like $E_{\tau}$, $E_{\tau, \pi}$ is monotonically decreasing in $\tau$, \ie, $\tau_1 > \tau_2$ implies $E_{\tau_1, \pi} \subset E_{\tau_2, \pi}$. Moreover, it is monotonically \emph{increasing} in $\pi$, \ie, $\pi_1 < \pi_2$ implies $E_{\tau, \pi_1} \subset E_{\tau, \pi_2}$, because a $\varTPA{\pi_1}$ is also a $\varTPA{\pi_2}$. For $\pi \leq 1$, a $\piTPA$ is empty, and $(\tau, \pi)$-compliance only requires $\tau$-sleepiness at all slots, \ie, $E_{\tau, \pi} = E_{\tau}$. The $(\tau, \pi)$-sleepy model is then indeed a generalization of the $\tau$-sleepy model. For $\pi = \infty$, a $\piTPA$ can be an unbounded period of asynchrony starting after slot $t_1$, and $(\tau, \pi)$-compliance is only defined for $\tau = \infty$ as well. It requires $|H_{t_1} \setminus A_{\infty}| > |A_{\infty} \cup (H_{0, \infty} \setminus H_{t_1})|$, where $A_{\infty}$ and $H_{0, \infty}$ are defined in the obvious way as limits. $H_{t_1} \setminus A_{\infty}$ are the honest voters of slot $t_1$ which are never corrupted, and $H_{0, \infty}$ are all honest validators which ever vote. If all validators vote at least once in the entire execution, then the requirement simply becomes $|H_{t_1} \setminus A_{\infty}| > \frac{n}{2}${, capturing the intuition that we can in principle get asynchronous safety as long as an honest majority of validators (which are never corrupted) agrees on something.}

\paragraph*{Aware validators} 
Given a $\piTPA (t_1, t_2)$, we say that a validator $v_i$ is \emph{aware at round $r$} for $r$ in slots $(t_1, t_2]$ if $v_i$ is active at round~$r$ \emph{and} $v_i \in H_{t_1}$. For $r$ not in slots $(t_1, t_2]$, we say that~$v_i$ is aware at round~$r$ simply if it is active at round~$r$. We motivate this notion after using it to define \emph{asynchrony resilience} (Definition~\ref{def:asynchrony-resilience}). 

\subsection{Security}
\paragraph*{Security Parameters}
We largely follow here the notation and definitions of~\cite{goldfish}.
We consider $\lambda$ and $\kappa$ be the security parameter associated with the cryptographic components used by the protocol and the security parameter of the protocol itself, respectively. We consider a finite time horizon $\Thorizon$, which is polynomial in $\kappa$. An event happens with \emph{overwhelming probability}, or w.o.p, if it happens except with probability which is $\negl(\kappa) + \negl(\lambda)$. Properties of cryptographic primitives hold except with probability $\negl(\lambda)$, \ie, with overwhelming probability, but we leave this implicit in the remainder of this~work. 

\paragraph*{Confirmed chain}
The protocols which we consider always specify a \emph{confirmation rule}, with whom validators can identify a \emph{confirmed prefix} of the canonical chain. Alongside the canonical chain, validators then also keep track of a \emph{confirmed chain}. We refer to the confirmed chain of validator~$v_i$ at round~$r$ as~$\chainconf_i^r$ (\cf $\chain_i^r$ for the canonical chain). This is the output of the protocol, for which safety properties should hold, and thus with respect to which the security of the protocol is defined. 

\begin{definition}[Secure protocol~\cite{goldfish}]
 \label{def:security}
We say that a consensus protocol outputting a confirmed chain $\chainconf$ is \emph{secure}, and has confirmation time $\Tconf$\footnote{If the protocol satisfies liveness, then at least one honest proposal is added to the confirmed chain of all active validators every $\Tconf$ slots. Since honest validators include all transactions they see, this ensures that transactions are confirmed within time $\Tconf + \Delta$ (assuming infinite block sizes or manageable transaction volume).}, if $\chainconf$ satisfies:
    \begin{itemize}
        \item \textbf{Safety:} For any two rounds $r, r'$, and any two honest validators $v_i$ and $v_j$ (possibly $i=j$) at rounds $r$ and $r'$ respectively, either $\chainconf_i^r \preceq \chainconf_{j}^{r'}$ or $\chainconf_j^{r'} \preceq \chainconf_i^r$.

        \item \textbf{Liveness:} For any rounds $r$ and $r' \geq r+\Tconf$, and any honest validator $v_i$ active at round $r'$, $\chainconf_{i}^{r'}$ contains a block proposed by an honest validator at a round $> r$.
    \end{itemize}
A protocol satisfies \emph{$\tau$-safety} and \emph{$\tau$-liveness} if it satisfies safety and liveness, respectively, \emph{in the $\tau$-sleepy model}, \ie, in $\tau$-compliant executions $E_{\tau}$. A protocol satisfies $\tau$-security if it satisfies $\tau$-safety and $\tau$-liveness.
\end{definition}

Observe that, for $\tau_1 > \tau_2$, since the $\tau_1$-sleepy model makes stronger assumptions than the $\tau_2$-sleepy model, security in the $\tau_1$-sleepy model is weaker than security in the $\tau_2$-sleepy model, \ie, \emph{$\tau_2$-security implies $\tau_1$-security}. This is immediate from $E_{\tau_1} \subset E_{\tau_2}$, because $\tau_2$-security is precisely security in all executions $E_{\tau_2}$, which implies security in $E_{\tau_1}$. 

\begin{definition}[Dynamic availability]
 \label{def:dyn-ava}
We say that a consensus protocol is $\tau$-\emph{dynamically-available} if and only if it satisfies $\tau$-security with confirmation time $\Tconf = O(\kappa)$. Moreover, we say that a protocol is dynamically available if it is $1$-dynamically-available, as this corresponds to the usual notion of dynamic availability.
\end{definition}

\begin{definition}[Reorg resilience]
\label{def:reorg-resilience}
An execution 
% in the partially synchronous network model
satisfies \emph{reorg resilience} if any honest proposal $B$ from a slot $t$ is always in the canonical chain of all active validators at rounds $\geq 3\Delta t + \Delta$.
A protocol is \emph{$\tau$-reorg-resilient} if all $\tau$-compliant executions satisfy reorg resilience.
\end{definition}

\begin{definition}[Asynchrony resilience]
\label{def:asynchrony-resilience}
An execution in the synchronous network model with a $\TPA \ (t_1, t_2)$ satisfies \emph{asynchrony resilience} if any honest proposal from a slot $t \leq t_1$ is always in the canonical chain of all \emph{aware} validators at rounds $\geq 3\Delta t + \Delta$.
A protocol is $(\tau, \pi)$-asynchrony-resilient if all $(\tau, \pi)$-compliant executions satisfy asynchrony resilience.
\end{definition}

In Definition~\ref{def:tau-pi-compliance}, we assume that validators $H_{t_1}$ are also awake at round $3\Delta t_1 + 2\Delta$, so that they observe their own votes, \ie, each validator in $H_{t_1}$ has all honest votes from slot $t_1$ in their view going forward. They are then the only validators which we can require to see all honest proposals from before the $\TPA$ as canonical \emph{during} the $\TPA$. For example, we cannot require a validator which is asleep at slot $t_1$, but active at slot $t_1 + 1$, to see an honest proposal from slot $t_1$ as canonical, because asynchrony has already started and they might not have received the proposal at all. After the $\TPA$, the requirement can again apply to all active validators. In other words, \emph{we define asynchrony resilience as reorg resilience of proposals made before the $\TPA$, in the views of aware validators}.

\section{Propose-vote-merge protocols}
\label{sec:pvm}
In this section we give a characterization of a class of protocols that we call \emph{propose-vote-merge} protocols. These are protocols that proceed in \emph{slots} consisting of $k$ rounds, each having a proposer $v_p$, chosen through a proposer selection mechanism, e.g., the one outlined in Section~\ref{sec:model}. In this work, unless otherwise specified, we analyze the case $k=3\Delta$.

At the beginning of each slot $t$, a block is proposed by $v_p$. All active validators (or \emph{voters}) vote after $\Delta$ rounds (what they vote for will become clear shortly). The last $\Delta$ rounds of the slot are needed for the \emph{view-merge} synchronization technique, as explained in Section~\ref{sec:view-merge}. Every validator $v_i$ has a buffer $\B_i$, a collection of messages received from other validators, and a view $\V_i$, used to make consensus decisions, which admits messages from the buffer only at specific points in time. 

Propose-vote-merge protocols are equipped with a deterministic fork-choice function $\FC$, which is used by honest proposers and voters to decide how to propose and vote, respectively, based on their view at the round in which they are performing those actions. It is moreover used as the basis of a \emph{confirmation rule}, as described in Section~\ref{sec:confirmation-rule}, with respect to which the security of the protocol is defined. We differentiate propose-vote-merge protocols based on which fork-choice they implement; in other words, these protocols are \emph{uniquely characterized by $\FC$}. In section~\ref{sec:pvm-properties} we prove properties about propose-vote-merge protocols in a fork-choice-agnostic way. Then, we will instead focus on protocols using $\ghost$-based fork-choice functions.

\subsection{Message types}
In propose-vote-merge protocols there are three message types, namely \textsc{propose}, \textsc{block}, and  \textsc{vote} messages. We make no distinctions between network-level representation of blocks and votes, and their representation in a validator's view, \ie, there is no difference between \textsc{block} and \textsc{vote} messages and blocks and votes, and we usually just refer to the latter. In the following description, $t$ is a slot and $v_i$ a validator. A block, or \textsc{block} message, is a tuple [\textsc{block}, $b$, $t$, $v_i$], where $b$ is a \emph{block body}, \ie, the protocol-specific content of the block\footnote{For simplicity, we omit a reference to the parent block. As mentioned in Section~\ref{sec:model}, we leave questions of validity implicit.}. A vote, or \textsc{vote} message, is a tuple [\textsc{vote}, $B$, $t$, $v_i$], where $B$ is a block. A proposal, or \textsc{propose} message, is a tuple [\textsc{propose}, $B$, $\V_i$, $t$, $v_i$] where $B$ is a block and $\V_i$ the view of validator $v_i$. Votes are gossiped at any time, and the same goes for blocks, regardless of whether they are received directly or as part of a vote or a proposal, \ie, a validator receiving a vote or proposal also gossips the block that it contains. Finally, a proposal from slot $t$ is gossiped only during the first $\Delta$ rounds of slot $t$.

\subsection{Protocol}
\label{sec:pvm-protocol}

\begin{figure}
    \centering
\begin{tikzpicture}
% Calculate segment width
\pgfmathsetmacro{\segwidth}{3.0} % increased segment width

% Draw horizontal line
\draw (0,0) -- (3*\segwidth,0);

% Draw vertical lines
\draw (0,-0.7) -- (0,0.7); % increased length
\draw (\segwidth,-0.35) -- (\segwidth,0.35); % increased length
\draw (2*\segwidth,-0.35) -- (2*\segwidth,0.35); % increased length
\draw (3*\segwidth,-0.7) -- (3*\segwidth,0.7); % increased length

% Add text labels
\node[above, align=center, text width=3cm] at (0,0.7) {\textbf{Propose} \\ Proposer merges view and buffer, broadcasts \textsc{propose} msg. based on it, containing a block and view};
\node[above, align=center, text width=3cm] at (\segwidth,0.35) {\textbf{Vote} \\ Validators merge the proposed view with theirs and broadcast \textsc{vote} msg. for the output of $\FC$};
\node[above, align=center, text width=3cm] at (2*\segwidth,0.35) {\textbf{Merge} \\ Validators merge buffer into view};
\node[above, align=center, text width=3cm] at (3*\segwidth,0.7) {\textbf{Propose}};

\node[below, align=center, text width=2.5cm] at (0,-0.7) {\textbf{$3\Delta t$}};
\node[below, align=center, text width=2.5cm] at (\segwidth,-0.35) {\textbf{$3\Delta t + \Delta$}};
\node[below, align=center, text width=2.5cm] at (2*\segwidth,-0.35) {\textbf{$3\Delta t + 2 \Delta$}};
\node[below, align=center, text width=2.5cm] at (3*\segwidth,-0.7) {\textbf{$3\Delta (t+1)$}};

\end{tikzpicture}
\vspace{-1.5em} % Adjusted the vertical spacing
\caption{Slot $t$ of a propose-vote-merge protocol, with its three phrases.}
\label{fig:pvm}
\end{figure}
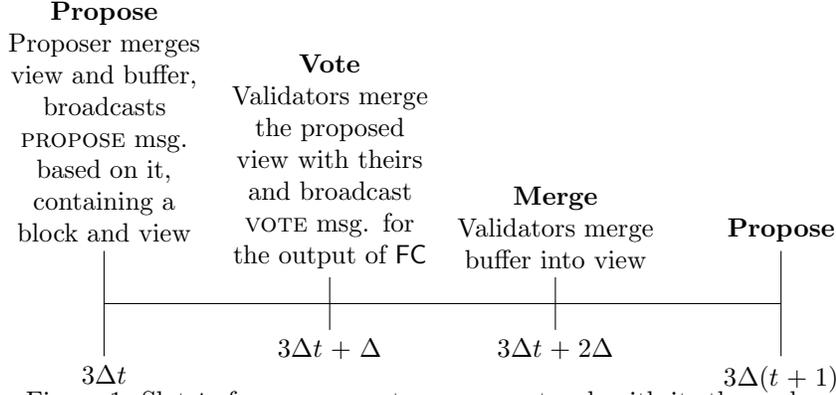

We now define a propose-vote-merge protocol with a generic fork-choice function $\FC$. Recall that a fork-choice function uniquely identifies a protocol of this family. A slot of the protocol proceeds in three phase, depicted in Figure~\ref{fig:pvm} and implemented in Algorithm~\ref{alg:pvm-generic}. Observe that validators (which have synchronized clocks) update the variables $t$ and $r$ representing slot and round, respectively, throughout the protocol's execution. 

\textsc{Propose}: At round $3\Delta t$, proposer $v_p$ merges its view $\V_p$ with its buffer $\B_p$, \ie, $\V_p \gets \V_p \cup \B_p$, and sets $\B_p \gets \emptyset$. Then,~$v_p$ runs the fork-choice function with inputs its view $\V_p$ and slot $t$, obtaining the head of the chain $B' = \FC(\V_p, t)$. Proposer $v_p$ extends $B'$ with a new block $B$, and updates its canonical chain accordingly, setting $\chain_p \gets B$. Finally, it broadcasts [\textsc{propose}, $B$, $\V_p\,\cup \{B\}$, $t$, $v_p$].

\textsc{Vote}: In rounds $[3\Delta, 3\Delta t + \Delta]$, every validator $v_i$ that receives a proposal message [\textsc{propose}, $B$, $\V$, $t$, $v_p$] from $v_p$ merges its view with the proposed view $\V$ by setting $\V_i \gets \V_i \cup \V$. At round $3\Delta t + \Delta$, regardless of whether or not $v_i$ received a proposal message, $v_i$ broadcasts the vote message [\textsc{vote}, $\FC(\V_i, t)$, $t$, $v_i$], and updates its canonical chain by setting $\chain_i \gets \FC(\V_i, t)$.

\textsc{Merge:} At round $3\Delta t + 2\Delta$, every validator $v_i$ merges its view with its buffer, \ie, $\V_i \gets \V_i \cup \B_i$, and sets $\B_i \gets \emptyset$. 

\begin{algo}
\caption{Propose-vote-merge protocol for $v_i$}
\label{alg:pvm-generic}
\vbox{
\small
\begin{numbertabbing}\reset
  xxxx\=xxxx\=xxxx\=xxxx\=xxxx\=xxxx\=MMMMMMMMMMMMMMMMMMM\=\kill
  \textbf{State} \label{}\\
  \> \(\V_i \gets \{\B_{\text{genesis}}\} \): view of validator $v_i$ \label{}\\
  \> \(\B_i \gets \emptyset \): buffer of validator $v_i$  \label{}\\
  \> \(\chain_i \gets B_{\text{genesis}}\): canonical chain of validator $v_i$ \label{}\\  
  \> \(t \gets 0\): the current slot \label{}\\
  \> \(r \gets 0\): the current round \label{}\\
  \textsc{propose}\\
  \textbf{at} $r=3\Delta t$ \textbf{do} \label{} \\
  \> \textbf{if} $v_i = v_p^t$ \textbf{then} \label{}\\
  \>\>$\V_i \gets \V_i \cup \B_i$ \label{}  \\
  \>\> $\B_i \gets \emptyset$ \label{}\\
  \>\> $B' \gets \FC(\V_i, t)$ \label{}\\
  \>\> $B \gets \mathsf{NewBlock}(B')$ \label{}\\
  \>\> // append a new block on top of $B'$ \label{} \\
  \>\> $\chain_i \gets B$ \label{}\\
  \>\> gossip message [\textsc{propose}, $B$, $\V_i\,\cup \{B\}$, $t$, $v_i$] \label{} \\
     \textsc{vote}\\
     \textbf{at} $r=3\Delta t + \Delta$ \textbf{do} \label{}\\
     \> $\chain_i \gets \FC(\V_i, t)$ \label{}\\
     \> gossip message [\textsc{vote}, $\FC(\V_i, t)$, $t$, $v_i$] \label{} \\
  \textsc{merge}\\
     \textbf{at} $r=3\Delta t + 2\Delta$ \textbf{do} \label{}\\
     \> $\V_i \gets \V_i \cup \B_i$ \label{}\\
       \> $\B_i \gets \emptyset$ \label{}\\
    \textbf{upon} receiving a message [\textsc{propose}, $B$, $\V$, $t$, $v_p^t$] \textbf{do} \label{}\\
    \> $\B_i \gets \B_i \cup \{B\}$ \label{}\\  
    \> \textbf{if} $r \in [3\Delta t, 3\Delta t + \Delta]$ \textbf{then} \label{}\\
    \>\> $\V_i \gets \V_i \cup \V$ \label{}\\
    \textbf{upon} receiving a message  [\textsc{vote}, $B$, $t'$, $v_i$] from \(v_i\) \textbf{do} \label{}\\
    \> $\B_i \gets \B_i \cup$ \{[\textsc{vote}, $B$, $t'$, $v_i$]\} \label{}\\  
    \textbf{upon} receiving a message [\textsc{block}, $b$, $t'$, $v_i$] from \(v_i\) \textbf{do} \label{}\\
    \> $\B_i \gets \B_i \cup $ \{[\textsc{block}, $b$, $t'$, $v_i$]\} \label{}
\end{numbertabbing}
}
\end{algo}

\subsection{View-merge}
\label{sec:view-merge}

The \textsc{merge} phase, along with all other operations involving views and buffers discussed in the previous section, are implementing the \emph{view-merge} technique. View-merge has been introduced by Kane, Fackler, Gagol, and Straszak~\cite{highway} to guarantee liveness of the Highway protocol, and then by D'Amato, Neu, Tas, and Tse~\cite{goldfish} to ensure \emph{reorg resilience}, \ie, proposals made by honest validators stay in the canonical chain, under synchrony. The core concept of the view-merge technique involves synchronizing the views of all honest validators with the view $\V_p$ of the proposer for a specific slot \emph{before} the validators broadcast their votes in that slot. To do so, view-merge works as follows:
\begin{enumerate}
    \item A validator's buffer is merged into its view only at one specific time, \ie, $\Delta$ rounds preceding the start of a new slot. Therefore, no new messages enter its view this way before voting in the next slot.
    \item The one exception to the above is the proposer of the next slot, which merges the buffer right before proposing, \emph{$\Delta$ rounds after all other validators}. It then proposes its resulting view $\V_p$ and a block extending the canonical chain as determined by $\FC$, according to $\V_p$.
    \item Validators merge the proposed view into their local view, execute $\FC$ based on this merged view, and vote.
\end{enumerate}

Observe that, if the network delay is less that $\Delta$ rounds, the view of the proposer is a superset of the views of other validators, because all messages merged by validators in the first step will also be merged by the proposer $\Delta$ rounds later. Then, the final merged view of all validators before voting is equal to the view of the proposer. If this is the case, since the output of the fork choice is a function of the view of a validator, every honest validator will have the same fork choice output, agreeing with the proposed block. As a consequence, active validators vote for honest proposals under synchrony. We refer to this as the \emph{view-merge property}, and prove it here for \emph{all propose-vote-merge protocols}. 

\begin{lemma}
\label{thm:view-merge-property}
Suppose that $t$ is a pivot slot. Then, all honest voters of slot $t$, \ie, $H_t$, vote for the honest proposal~$B$ of slot~$t$.
\end{lemma}

\begin{proof}
Let $\V_p \cup \{B\}$ be the view proposed with block $B$ by~$v_p$, the honest proposer of slot $t$, \ie, $\V_p$ is the view of~$v_p$ at round $3\Delta t$. Since $v_p$ is honest, $B$ extends $\FC(\V_p, t)$, and thus $\FC(\V_p \cup \{B\}, t) = B$ by the consistency property (see Section~\ref{sec:model}) of~$\FC$.

Consider an honest voter of slot $t$, \ie, a validator $v_i \in H_t$, and let $\V_i$ be its view at round $3\Delta t + \Delta$, before merging $\V_i$ with the proposed view $\V_p \cup \{B\}$.
Observe that, since $v_i$ is active in round $3\Delta t + \Delta$, it must has already been awake at round $3\Delta (t-1) - 2\Delta$, because otherwise it would need to follow the joining protocol until round $3\Delta t + 2\Delta$, and would thus not currently be active. 

Therefore, $v_i$ was already active at round $3\Delta (t-1) - 2\Delta$, and in particular it merged its buffer $\B_i$ in its local view then. So, $\V_i$ is the view that $v_i$ had after merging the buffer $\B_i$. So, messages in $\V_i$ are delivered to the proposer by round $3\Delta t$, so $\V_i \subseteq \V_p$.

The proposal message is received by $v_i$ before voting. Then, $v_i$ merges the proposed view $\V_p \cup \{B\}$ with its view $\V_i$, resulting in the view $\V_i \cup (\V_p \cup \{B\}) = \V_p \cup \{B\}$. 
Validator $v_i$ votes for the output of its fork-choice at round $3\Delta t + \Delta$, which is $\FC(\V_p \cup \{B\}, t) = B$.
\end{proof}

\subsection{Joining protocol}
\label{sec:joining-protocol}

In Section~\ref{sec:model} we described a model in which honest validators that become awake at round $r$, before starting to participate in the protocol, must first execute (and terminate) a \emph{joining protocol}, after which they become \emph{active}. We now recall such a protocol, as presented in~\cite{goldfish}.

When an honest validator $v_i$ wakes up at some round $r \in (3\Delta (t-1) + 2\Delta, 3\Delta t + 2\Delta]$, it immediately receives all the messages that were sent while it was asleep, and it adds them into its buffer $\B_i$, without actively participating in the protocol yet. All new messages which are received are added to the buffer $\B_i$, as usual. Validator~$v_i$ then waits for the \emph{next view-merge opportunity}, at round $3\Delta t + 2\Delta$, in order to merge its buffer $B_i$ into its view $V_i$. At this point, $v_i$ starts executing the protocol. From this point on, validator $v_i$ becomes \emph{active}, until either corrupted or put to sleep by the adversary.

\subsection{Confirmation rule}
\label{sec:confirmation-rule}

In propose-vote-merge protocols, the confirmed chain $\chainconf$ is the $\kappa$-deep prefix \emph{in terms of slots} of the canonical chain $\chain$, \ie, its prefix corresponding to blocks proposed at slots $\leq t - \kappa$, which we denote with $\chain^{\lceil \kappa}$. A validator $v_i$ updates its confirmed chain $\chainconf_i$ whenever it updates its canonical chain $\chain_i$ by computing the fork-choice, \ie, at round $3\Delta t + \Delta$, and possibly also $3\Delta t$, if they are the proposer of slot $t$, so that at any time we have $\chainconf_i = \chain_i^{\lceil \kappa}$. We show that, with overwhelming probability, all intervals $[t-\kappa, t)$ of $\kappa$ consecutive slots contain a \emph{pivot slot}, \ie, a slot with an honest proposer which is active at proposal time, and thus which makes a proposal. To achieve this, we rely on the lower bound $h_0$ on the active validators $h_r$ at any round, which guarantees that in any slot there is at least a probability $\frac{h_0}{n}$ of an honest proposal being made.

\begin{lemma}
\label{thm:frequent-honest-proposals-whp}
With overwhelming probability, all slot intervals of length $\kappa$ contain at least a pivot slot. 
\end{lemma}

\begin{proof}
By assumption of \emph{fairness} of the proposal mechanism, the proposer $v_p$ of slot $t$ is active at round $3\Delta t$ with probability $\frac{h_{3\Delta t}}{n} \geq \frac{h_0}{n}$, for $h_0 > 0$. Given any $\kappa$ slots, the probability of none of the $\kappa$ slots having an active proposer is $\leq (\frac{n - h_0}{n})^\kappa$, \ie, negligible in~$\kappa$. The number of slot intervals of length $\kappa$ which we need to consider is equal to the time horizon $\Thorizon$ over which the protocol is executed, which is polynomial in~$\kappa$, so the probability of even one occurrence of~$\kappa$ consecutive slots without a pivot slot is also negligible.
\end{proof}

\subsection{Properties}
\label{sec:pvm-properties}

We now prove properties that propose-vote-merge protocols with a generic fork-choice function $\FC$ satisfy. An important property is reorg resilience (Definition~\ref{def:reorg-resilience}), which we show implies security (Definition~\ref{def:security}) in Theorem~\ref{thm:pvm-dynamic-availability}. The two key ingredients for reorg resilience are the view-merge property (Lemma~\ref{thm:view-merge-property}), and the following proposition. While the former holds for any propose-vote-merge protocol, the latter does only for some of them, and only in some $\tau$-sleepy model. In particular, we later prove that it holds for $\FC = \ghosteph$, the fork-choice of \Goldfish~\cite{goldfish}, if $1$-sleepiness is satisfied. More generally, we later show that it holds for $\FC = \ulmdghost$ with vote expiry parameter $\eta$, \ie, the fork-choice rule introduced in this work and presented in Section~\ref{sec:ULMD}, if $\eta$-sleepiness is satisfied. 

\begin{proposition}
\label{prop:induction-step-assumption}
Suppose that all honest voters of slot $t-1$ vote for a descendant of block $B$. Then, $B$ is in the canonical chain of all active validators in rounds $\{3\Delta t, 3\Delta t + \Delta\}$. In particular, all honest voters of slot $t$ vote for descendants of~$B$.
\end{proposition}

We now show that, if Proposition~\ref{prop:induction-step-assumption} holds for an execution, then the execution satisfies reorg resilience. The idea is the following: by the view-merge property (Lemma~\ref{thm:view-merge-property}), all active validators vote for honest proposals, and Proposition~\ref{prop:induction-step-assumption} ensures that this keeps holding also in future slots. We prove this result in the following theorem, which immediately implies that a protocol is $\tau$-reorg-resilient if Proposition~\ref{prop:induction-step-assumption} holds for it in the $\tau$-sleepy model.

\begin{theorem}[Reorg resilience]
\label{thm:pvm-reorg-resilience}
Let us consider an execution of a propose-vote-merge protocol in which Proposition~\ref{prop:induction-step-assumption} holds. Then, this execution satisfies reorg resilience.
\end{theorem}

\begin{proof}
Consider a honest proposal $B$ from slot $t$. We prove reorg resilience by induction on the slot. Note that validators only ever update their canonical chain at rounds $\{3\Delta s, 3\Delta s + \Delta\}$, for all slots $s \geq t$, upon computing the fork-choice. Therefore, the following statement holding for all $s \geq t$ is sufficient for reorg resilience, as it implies that $B$ is canonical in all rounds $\geq 3\Delta t + \Delta$. 

\textbf{Induction hypothesis:} $B$ is canonical in the views of active validators at rounds $r \in \{3\Delta s, 3\Delta s + \Delta\}$, for a slot $s\geq t$ and $r \geq 3\Delta t + \Delta$. 

\textbf{Base case:} The proposal slot $t$. Lemma~\ref{thm:view-merge-property} applies and implies that all honest voters at slot $t$ vote for $B$, which is in particular canonical in their views.

\textbf{Inductive step:} Suppose now that the statement holds for $s \geq t$. In particular, all honest voters of slot~$s$ vote for a descendant of~$B$, because it is canonical in their view in the voting round $3\Delta s + \Delta$. Proposition~\ref{prop:induction-step-assumption} then implies the desired statement for $s+1$.
\end{proof}

% Observe that, though we later prove Proposition~\ref{prop:induction-step-assumption} only in the context of $\FC = \ghosteph$ under sleepiness and $\FC = \ulmdghost$ under $(\eta-1)$-sleepiness, this does not exclude the possibility of obtaining other fork-choice functions that satisfy Proposition~\ref{prop:induction-step-assumption} under some assumptions, and which are thus also reorg resilient under those assumptions.

If an execution satisfies reorg resilience we obtain that, by applying the same arguments as in~\cite{goldfish}, the $\kappa$-deep confirmation rule is secure in it, in the sense that the confirmed chain satisfies Definition~\ref{def:security}. In particular, $\tau$-reorg resilience implies $\tau$-dynamic-availability. Because of Thereom~\ref{thm:pvm-reorg-resilience}, we then only need to show that Proposition~\ref{prop:induction-step-assumption} holds for $\tau$-compliant executions in order to show that a protocol is $\tau$-dynamically-available.

\begin{theorem}[Dynamic-availability]
\label{thm:pvm-dynamic-availability}
An execution of a propose-vote-merge protocol satisfying reorg resilience also satisfies security with overwhelming probability with $\Tconf = 2\kappa$ slots. In particular, $\tau$-reorg-resilience implies $\tau$-dynamic-availability.
\end{theorem}

\begin{proof}
Theorem~\ref{thm:pvm-reorg-resilience} and Lemma~\ref{thm:frequent-honest-proposals-whp} imply security (Definition~\ref{def:security}) with overwhelming probability, as we now explain. For a round~$r$, denote by $\text{slot}(r)$ the slot to which that round belongs. We show liveness with confirmation time $\Tconf = 2\kappa$ slots. Consider a round $r$, with $t = \text{slot}(r)$, a round $r'$ with $t' = \text{slot}(r') \geq t + 2\kappa$, and an honest validator $v_i$ active at round $r'$. By Lemma~\ref{thm:frequent-honest-proposals-whp}, w.o.p, there exists a pivot slot $t'' \in [t+1, t+\kappa]$. By Theorem~\ref{thm:pvm-reorg-resilience}, the proposal $B$ from slot $t''$ is in the canonical chain of all active validators in later slots, so in particular it is in $\chain_i^{r'}$. Since $t'' \leq t+\kappa \leq t' - \kappa$, $B$ is $\kappa$-deep in $\chain_i^{r'}$, and so it is in the confirmed chain $\chainconf_i^{r'}$ as well.

To show safety, let us consider any two rounds $r' \ge r$, and any two honest validators $v_i$ and $v_j$, active at rounds $r$ and $r'$, respectively. Let also $t = \text{slot}(r)$. Lemma~\ref{thm:frequent-honest-proposals-whp} implies that w.o.p. there is at least a pivot slot $t' \in [t-\kappa, t)$, and by Theorem~\ref{thm:pvm-reorg-resilience} its proposal $B$ is canonical in all active views from round $3\Delta t' + \Delta$. Therefore,~$B$ is in the canonical chain of $v_i$ at round $r$ and, since it is from a slot $\geq t-\kappa$, $\chainconf_i^r \preceq B$. Block $B$ is also in the canonical chain of $v_j$ at round $r'$, \ie, either $B \preceq \chainconf^{r'}_j$ or $\chainconf^{r'}_j \preceq B$. In the first case, $\chainconf_i^r \preceq B \preceq \chainconf^{r'}_j$. In the second case, we have both $\chainconf_i^r \preceq B$ and $\chainconf^{r'}_j \preceq B$. Therefore, $\chainconf_i^r$ and $\chainconf^{r'}_j$ cannot be conflicting, it follows that either $\chainconf_{i}^{r} \preceq \chainconf_{j}^{r'}$ or $\chainconf_{j}^{r'} \preceq \chainconf_{i}^{r}$.
\end{proof}

\section{Propose-vote-merge protocols based on GHOST}

\subsection{Prerequisites}
\label{sec:prereqs}
 
In this section we recall the $\lmdghost$ fork-choice function as utilized in the Ethereum consensus protocol, Gasper~\cite{gasper}. We start by presenting the fork-choice function $\ghost$, the main building block of $\lmdghost$, and of all the fork-choice functions we consider in this work. %Limitations of \LMDGHOST\footnote{Note that $\lmdghost$ denotes the fork-choice function, whereas \LMDGHOST refers to the dynamically available protocol implemented in Gasper~\cite{gasper}.} can be found in Appendix~\ref{sec:basic-lmd-ghost-limitations}.

\subsubsection{GHOST}
\label{sec:ghost-fork-choice}

$\ghost$ is a fork-choice function based on the fork-choice procedure introduced in~\cite{ghost} by Sompolinsky and Zohar, a greedy algorithm that grows a blockchain on sub-branches with the \emph{most activity}. Except, this one is \emph{vote-based} rather than \emph{block-based}, \ie, here we weigh sub-trees based on number of votes rather than blocks. Given a set of votes $M$, we define the \emph{weight} function $w(B,M)$ to output the number of votes in $M$ for $B$ or descendants of $B$, \ie, on the sub-tree rooted at $B$. Starting at the first block of the canonical chain, \ie, $B_{\text{genesis}}$, and considering the set $M$ of votes in $\V$, $\ghost$ iterates over a sequence of blocks from $\V$, selecting as the next block the descendant of the current block with the highest weight. This continues until it reaches a block that does not have any descendant in $\V$, which is output. We now state and prove a simple property of the $\ghost$ fork-choice, which we are going to repeatedly use throughout the work,  without explicitly mentioning it, whenever wanting to prove that a block is in the canonical chain in some view.

\begin{lemma}
\label{lem:ghost-majority-implies-canonical}
Let $\V$ be a view in which over a majority of the votes are for a descendant of a block $B$. Then, $\ghost(\V, t)$ is a descendant of $B$, \ie, $B$ is in the canonical chain output by the $\ghost$ fork-choice.
\end{lemma}

\begin{proof}
Let $M$ be all votes in $\V$. Consider any height less than or equal to the height of $B$. In any fork at such a height, there is one branch that contains $B$, and thus also the whole sub-tree rooted at $B$. Say the block on that branch at that height is $B'$, and consider any competing sibling $B''$. Since over a majority of the votes in $M$ are for the sub-tree rooted at $B$, and all votes on the sub-tree rooted at $B'$ are not votes on the sub-tree rooted at $B''$, $w(B', M) > \frac{|M|}{2} > w(B'',M)$. Thus, $B'$ is selected by the $\ghost$ fork-choice algorithm at that height. Therefore, $B\preceq \ghost(\V, t)$.
\end{proof}

\begin{algo}
\caption{$\ghost$ Fork-Choice function}
\label{alg:ghost}
\vbox{
\small
\begin{numbertabbing}\reset
  xxxx\=xxxx\=xxxx\=xxxx\=xxxx\=xxxx\=MMMMMMMMMMMMMMMMMMM\=\kill
  \textbf{function} $\ghost(\V, t)$ \label{}\\
  \> $B \gets B_{\text{genesis}}$ \label{}\\
  \> $M \gets $ all votes in $\V$\label{}\\
  \> \textbf{while} $B$ has descendant blocks in $\V$ \textbf{do} \label{}\\
  \>\> $B \gets \underset{B'\in \V, \text{ child of } B}{\text{arg max}} w(B',M)$ \label{} \\
  \>\> // ties are broken according to a deterministic rule \label{}\\
  \> \textbf{return} $B$ \label{}
\end{numbertabbing}
}
\end{algo}

\subsubsection{Filtered GHOST}
\label{sec:ghost-based-fc}

We define the family of $\ghost$-based fork-choice functions $\filteredghost$. \sloppy{A fork-choice function $\FC \in \filteredghost$ is characterized by a view filter $\FIL$, which takes as input a view $\V$ and a slot $t$, and outputs $(\V', t)$, where $\V'$ is another view such that $\V' \subseteq \V$. Then, $\FC(\V, t) \coloneqq \ghost(\FIL(\V, t))$, \ie, $\FC \coloneqq \ghost \circ\,\FIL$.} $\ghost$ itself is contained in $\filteredghost$, characterized by the identity filter. %Moreover, both $\lmdghost$ and $\ghosteph$, the fork-choice function of the \Goldfish protocol~\cite{goldfish} (presented in Section~\ref{sec:goldfish}) belong to $\filteredghost$: the former is characterized by a filter that removes all but the \emph{latest} votes (see Section~\ref{sec:lmd-ghost}), and the latter by a filter $\FIL(\V, t)$ that removes all the \emph{expired} votes from $\V$, \ie, from slots $< t-1$ (see Section~\ref{sec:vote-expiry}). As we see in Section~\ref{sec:ULMD}, both $\lmdghost$ and $\ghosteph$ are special cases of the $\ulmdghost$ fork-choice, which is also a member of $\filteredghost$.

\paragraph*{Equivocation discounting}
All fork-choice functions we consider from now on implement a technique to deal with equivocations, used in the Ethereum protocol, as a spam-resistance measure, \ie, \emph{equivocation discounting}~\cite{equiv-disc}\cite{goldfish}. It consists of excluding votes from equivocating validators from one's view, before running the fork-choice function on it. We describe equivocation discounting using view filters. Consider the view filter $\FIL_{eq}$ such that $\FIL_{eq}(\V, t)$ removes all votes by \emph{equivocating validators in $\V$}, \ie, validators for which $\V$ contains multiple, equivocating, votes for some slot $t$. Given a fork-choice function $\FC \in \filteredghost$, characterized by the view filter $\FIL$, we apply equivocation discounting to $\FC$ by composing $\FIL$ with $\FIL_{eq}$, \ie, by considering the fork-choice function $\FC_{eq} \coloneqq \ghost\circ\,\FIL\circ\,\FIL_{eq}$. All fork-choice functions considered in the following sections implement equivocation discounting, even if not explicitly stated.

\subsubsection{LMD-GHOST}
\label{sec:lmd}
$\lmdghost$ is an adaptation of the original $\ghost$~\cite{ghost}, in which the protocol considers only each validator’s most recent vote (LMD), which is assumed to be the most meaningful. $\lmdghost$ belongs to the class $\filteredghost$ of fork-choice functions based on $\ghost$~\cite{ghost}, and is characterized by a view filter $\FIL_{\text{lmd}}$. $\FIL_{\text{lmd}}(\V, t)$ removes all but the latest votes of every validator (possibly more than one) from $\V$ and outputs the resulting view, \ie, it implements the \emph{latest message} (LMD) rule. 

 \begin{algo}
 \caption{$\lmdghost$ Fork-Choice function}
 \label{alg:lmd-ghost}
 \vbox{
 \small
 \begin{numbertabbing}\reset
   xxxx\=xxxx\=xxxx\=xxxx\=xxxx\=xxxx\=MMMMMMMMMMMMMMMMMMM\=\kill
   \textbf{function} $\lmdghost(\V, t)$ \label{}\\
   \> \textbf{return} $\ghost(\FIL_{\text{lmd}}(\FIL_{\text{eq}}(\V, t)))$ \label{}\\
   \textbf{function} $\FIL_{\text{lmd}}(\V, t)$ \label{}\\
   \> $\V' \gets \V$ without all but the most recent (\emph{latest}) votes \label{} \\
   \>\>\> of each validator \label{}\\
   \> \textbf{return} $(\V', t)$ \label{}\\
   \textbf{function} $\FIL_{\text{eq}}(\V, t)$ \label{}\\
   \> $\V' \gets \V$ without all votes by equivocators in $\V$ \label{}\\
   \> \textbf{return} $(\V', t)$ \label{}
 \end{numbertabbing}
 }
 \end{algo}

\subsection{\LMDGHOST with view-merge and no subsampling}
\label{sec:lmd-ghost}

The first propose-vote-merge protocol that we consider is a variation of the original \LMDGHOST protocol, with the addition of view-merge and without considering subsampling of validators, the latter in order to prevent \emph{ex-ante reorgs}\footnote{The Ethereum protocol has (disjoint) committees of (the whole set of) validators voting in each slot, \ie, it implements \emph{subsampling}. Neither proposer boost nor view-merge can fully prevent ex-ante reorgs in that setting, leading to a protocol with different security guarantees than what is described here. In particular the \LMDGHOST protocol implemented in Ethereum is not a reorg resilient protocol, even in the full participation setting.}.

To specify a propose-vote-merge protocol, we only need to define the fork-choice function which uniquely characterizes it, which in this case is of course $\lmdghost$, as introduced in Section~\ref{sec:lmd}. From now on, we refer to this propose-vote-merge protocol simply as \LMDGHOST. 

{We first introduce the following notation. We denote with $\Hinf_t$ the set of honest voters of slot $t$ that \emph{are never corrupted}, \ie, $\Hinf_t \coloneqq \lim_{s \to \infty} H_t \setminus A_{s} = H_t \setminus A_{\infty}$. We refer to validators that are never corrupted as \emph{permanently honest}, and to $\Hinf_t$ as the \emph{permanently honest voters of slot $t$}.}

In the following theorem we analyze the reorg resilience of \LMDGHOST. In particular, given $|\Hinf_t| > \frac{n}{2}$ and an honest proposal $B$ from a slot $t$, then $B$ is always canonical in all honest views that contain all slot $t$ votes from $\Hinf_t$, \emph{without requiring synchrony at any future slot}. In other words, honest proposals made during synchrony immediately become \emph{asynchronously safe}. This is much stronger than the usual notion of reorg resilience, which requires \emph{continued synchrony}.

\begin{theorem}[Strong reorg resilience]
\label{thm:lmd-ghost-strong-reorg-resilience}
Consider an honest proposal $B$ from a slot $t$ in which network synchrony hold and $|\Hinf_t| > \frac{n}{2}$. Suppose that validators in $\Hinf_t$ do not fall asleep in rounds $[3\Delta t + \Delta, 3\Delta t + 2\Delta]$. Then, $B$ is always canonical in all honest views which contain all slot $t$ votes from $\Hinf_t$.
\end{theorem}

\begin{proof}
By Lemma~\ref{thm:view-merge-property}, all honest voters of slot~$t$ broadcast a vote for~$B$ at round $3\Delta t + \Delta$. Synchrony in the subsequent $\Delta$ rounds means that all such votes are received by those same validators before they merge their buffers, since by assumption they do not fall asleep. Those votes are then in all of their views by the end of slot~$t$ and the result follows.
\end{proof}

On the other hand \LMDGHOST is significantly limited in its support of dynamic participation, as shown in the following theorem. In particular, we present a scenario in which the adversary is able to cause a reorg of a confirmed block, compromising $\tau$-safety and, consequently, $\tau$-dynamic-availability, while never violating $\tau$-sleepiness. The reason why this attack is possible is due to the fact that $\tau$-sleepiness only considers votes from the last $\tau$ slots, but \LMDGHOST does not have vote expiry, so all votes are relevant to the fork-choice. Since the $\infty$-sleepy model allows only an extremely restrictive form of dynamic participation, almost equivalent to requiring $|\Hinf_t| > \frac{n}{2}$ at all times, this is a fairly strong limitative result.

\begin{theorem}
\label{thm:lmd-not-da}
\LMDGHOST is not $\tau$-dynamically-available for any finite $\tau$ and any confirmation rule with finite confirmation time $\Tconf$.
\end{theorem}

\begin{proof}
For some $\tau < \infty$ and a confirmation rule with confirmation time $\Tconf$, we show that $\tau$-safety and $\tau$-liveness are in conflict for \LMDGHOST. We look at a specific execution, which we assume satisfies liveness, and show that it does not satisfy safety. Moreover, we show that such execution is $\tau$-compliant. Therefore, there are $\tau$-compliant executions in which either liveness or safety is not satisfied, and consequently \LMDGHOST is not $\tau$-dynamically-available.

Without loss of generality, we fix a finite $\tau \geq \Tconf$ (we do not need to consider $\tau < \Tconf$ since $\tau_1$-dynamic-availability implies $\tau_2$-dynamic-availability for $\tau_1 \leq \tau_2$). We consider a validator set of size $n = 2m+1$, partitioned in three sets, $V_1$, $V_2$, and $V_3$, with $V_1 = \{v_1\}$, $|V_2| = m+1$, $|V_3| = m-1$. Validators in $V_2$ and $V_3$ are all initially honest, while $v_1$ is adversarial. Let~$t-1$ and~$t$ be two adversarial slots, \ie, controlled by $v_1$. In slot~$t$, validator $v_1$ publishes conflicting blocks $A$ and $B$, one as a proposal for slot $t-1$ and the other for slot~$t$. By round $3\Delta t + \Delta$, the adversary delivers only $A$ to validators in $V_2$, and only $B$ to validators in $V_3$, so that the former vote for $A$ and the latter for $B$ in slot~$t$\footnote{In practice, such splitting of honest views can be done without targeting specific validators, and rather only aiming to get an approximate partition of the honest validators in two. This kind of attack method has been discussed at length in the context of balancing attacks~\cite{ethresearch-balancing-attack2}.}. At this point, the adversary puts all validators in $V_3$ to sleep, and then does nothing for $N \gg \tau$ slots, \ie, until slot $t+N$. Meanwhile, validators in $V_2$ keep voting for $A$, since $V_2$ contains $m+1 > \frac{n}{2}$ validators, so $A$ stays canonical in all of the views of every member of $V_2$. Since $\tau \geq \Tconf$, this execution satisfying liveness implies that some honest proposal made after slot $t$ is confirmed in this period, and thus that block $A$ is confirmed, since all honest proposals made in this period are descendants of $A$. For any slot $s \in [0, t+1]$, we have that $|H_{s-1}| = |V_2 \cup V_3| = 2m$, so $\tau$-sleepiness is satisfied. For $s \in [t+2, t+\tau]$, we have that $|H_{s-1}| = |V_2| = m+1 > m = |V_1 \cup V_3| = |A_{s} \cup (H_{s-\tau, s-2}\setminus H_{s-1})|$, so $\tau$-sleepiness is also satisfied. For $s \in [t + \tau + 1, t+N-1]$, the first two terms are unchanged, while $H_{s-\tau, s-2} \setminus H_{s} = \emptyset$, because the last vote broadcast by the validators in $|V_3|$ is from slot $t < s - \tau$. $\tau$-sleepiness is then still satisfied. At slot $t+N$, the adversary corrupts a single validator $v_2 \in V_2$, and starts voting for $B$ with both $v_1$ and $v_2$. Now, $B$ has $m+1$ votes, and descendants of $A$ only $m$, so $B$ becomes canonical and stay so. After $\Tconf$ slots, it is confirmed by all validators in $V_2$, meaning we have a safety violation. The adversary does not perform any more corruptions nor puts to sleep any more validators, and does not wake up validators in $V_3$. Therefore, for all slots $s \geq t+N$, we have $A_{s} = \{v_1, v_2\}$, $V_2 \setminus \{v_2\} \subseteq H_{s-1}$ and $H_{s-\tau, s-2}\setminus H_{s-1} = \emptyset$. $\tau$-sleepiness is then satisfied, because $|H_{s-1}| \geq m > 2 = |A_{s} \cup H_{s-\tau, s-2}\setminus H_{s}|$. Therefore, the executions is $\tau$-compliant, and thus the protocol does not satisfy $\tau$-security.
\end{proof}

\subsection{\Goldfish}
\label{sec:goldfish}

\Goldfish is a simplified variant of \LMDGHOST, introduced by D’Amato, Neu, Tas, and Tse~\cite{goldfish}, which very nearly belongs to the family of propose-vote-merge protocols. Goldfish can tolerate dynamic participation, it supports subsampling of validators, and it is $1$-reorg-resilient and $1$-dynamically-available in synchronous networks with dynamic participation. During each slot in \Goldfish, only votes from the immediately preceding slot influence the protocol’s behavior.

In the following analysis, we depart slightly from the original formulation~\cite{goldfish}, and consider a version of \Goldfish which fulfills the specification of a propose-view-merge protocol. This variant does not consider subsampling and replaces the VRF lottery considered in the original protocol with an SSLE proposer selection mechanism (see Section~\ref{sec:model}). Note that accommodating these features of the original \Goldfish protocol would only require a small extension of the family of propose-vote-merge protocols, which we do not consider here for simplicity. In fact, the analysis of~\cite{goldfish} is almost exactly identical to the analysis of this variant of \Goldfish, in particular relying on reorg resilience in the same way. Crucially, in the case of \Goldfish, Proposition~\ref{prop:induction-step-assumption} is true even if subsampling is considered, due to the strict vote expiry.

% The confirmation rule is unchanged, but a consequence of having a proposer selection mechanism satisfying uniqueness is that a longer confirmation time, \ie, a higher~$\kappa$, is required to ensure the same failure probability.

\Goldfish is characterized by $\ghosteph$, a fork-choice function in $\filteredghost$ that takes a view $\V$ and a slot $t$ as inputs, and finds the canonical chain determined by the votes within $\V$ that were broadcast for slot~$t-1$.

\subsubsection{Vote Expiry}
\label{sec:vote-expiry}

The notion of \emph{vote expiry} has been introduced in the context of \Goldfish, where all but the \emph{most recent} votes are discarded. We generalize here this notion by introducing an \emph{expiration period} $\eta$. In particular, given a slot $t$ and a constant $\eta \in \mathbb{N}$ with $\eta \ge 0$, the \emph{expiration period} for slot $t$ is the interval $[t-\eta, t)$, and only votes broadcast within this period influence the protocol's behavior at slot~$t$. In particular, by assuming an expiration period, the capabilities of the adversary are limited, as votes broadcast before slot $t-\eta$ are not considered.

Formally, we can define the \emph{$\ghost$ fork-choice function with expiry period $\eta$} as a fork-choice function in $\filteredghost$. It is characterized by the filter function $\FIL_{\eta\text{-exp}}(\V, t)$ which removes all votes from slots $< t-\eta$ from $\V$, and outputs the resulting view. For $\eta = \infty$, we get back the regular $\ghost$ fork-choice function, whereas for $\eta = 1$, we get $\ghosteph$.

 \begin{algo}
 \vbox{
 \small
 \begin{numbertabbing}\reset
   xxxx\=xxxx\=xxxx\=xxxx\=xxxx\=xxxx\=MMMMMMMMMMMMMMMMMMM\=\kill
   \textbf{function} $\ghosteph(\V,t)$ \label{}\\
   \> \textbf{return} $\ghost(\FIL_{1\text{-exp}}(\FIL_{\text{eq}}(\V, t)))$ \label{}\\
     \\
   \textbf{function} $\FIL_{\eta\text{-exp}}(\V, t)$ \label{}\\
   \> $\V' \gets \V$ without all votes from slots $ < t-\eta$ \label{} \\
   \> \textbf{return} $(\V', t)$ \label{}
 \end{numbertabbing}
 }
 \caption{$\ghosteph$ Fork-Choice function}
 \end{algo}

Vote expiry allows the protocol to support dynamic participation. 

\subsubsection{Properties and limitations}

D'Amato, Neu, Tas, and Tse~\cite{goldfish} show that \Goldfish is $1$-reorg-resilient and $1$-dynamic-available, as also follows from Theorems~\ref{thm:ulmd-reorg-resilience} and~\ref{thm:ulmd-dynamic-availability} in our analysis of \ULMDGHOST, for the special case $\eta = 1$. On the other hand, \Goldfish is brittle to temporary asynchrony. Due to the expiry period being $\eta = 1$, even a single violation of the bound of $\Delta$ rounds on the network delay can lead to a catastrophic failure, jeopardizing the safety of \emph{any} previously confirmed block. This holds even with a single adversarial validator, and with all validators awake. Suppose for example that network delay is $ >\Delta$ rounds between rounds $[3\Delta t + \Delta, 3\Delta t + 2\Delta]$. This causes all honest votes broadcast at round $3\Delta t + \Delta$ to not be delivered to any honest validator by round $3\Delta t + 2\Delta$. Then, such votes are not in any honest view after merging the buffer. Suppose also that the proposer of slot~$t+1$ is malicious and proposes a block $B$ extending a block~$A$ which is not in the canonical chain. Moreover, the proposed view contains no votes other than a single slot~$t$ vote for~$A$ from~$v_p^{t+1}$ itself. Then, the view of an honest validator at the voting round $3\Delta (t+1) + \Delta$ contains only two slot~$t$ votes: its own, and the one for~$A$. If~$A$ wins the tiebreaker, all honest validators thus vote for~$B$, making it canonical and reorging all previously confirmed blocks. Since the period of asynchrony lasts for (less than) a single slot, all validators are always awake and there is a single adversarial validator, it is clear that the described execution is~$(\infty, 2)$-compliant. Since $E_{\tau, \pi}$ is monotonically decreasing in~$\tau$ and increasing in~$\pi$, and thus $E_{\infty, 2} \subseteq E_{\tau, \pi}$ for all $\tau > \pi \geq 2$, the above amounts to a proof of the following theorem.

\begin{theorem}
\label{thm:goldfish-not-asynchrony-resilient}
\Goldfish is not $(\tau, \pi)$-asynchrony-resilient for any $\tau > \pi \geq 2$.
\end{theorem}

\section{Recent Latest Message Driven (RLMD) GHOST}
\label{sec:ULMD}

In this section we present our propose-vote-merge protocol, \ULMDGHOST, which generalizes both \LMDGHOST and \Goldfish. It is characterized by the $\ghost$-based fork-choice $\FC = \ulmdghost$, which combines vote expiry (see Section~\ref{sec:vote-expiry}) with the latest message driven (LMD) fork-choice (see Section~\ref{sec:lmd}). Its filter function $\FIL_{\text{rlmd}}(\V, t)$ removes \emph{all but the latest messages within the expiry period $[t-\eta, t)$ for slot~$t$}, \ie, $\FIL_{\text{rlmd}} = \FIL_{\text{lmd}}\circ\FIL_{\eta\text{-exp}}\circ\FIL_{eq}$.

For $\eta = 1$, $\ulmdghost$ coincides with $\ghosteph$, and thus \ULMDGHOST with \Goldfish, because $\FIL_{1\text{-exp}}$ only considers votes from slot $t-1$, which are a subset of the latest votes, so that the LMD rule does not add any further filtering. For $\eta = \infty$, it coincides with $\lmdghost$, because no messages expire, so $\FIL_{\infty\text{-exp}}$ does not perform any filtering. Unless specified, we henceforth refer to $\ulmdghost$ with a generic parameter $\eta$. 

As \Goldfish, \ULMDGHOST can support \emph{fast confirmations} of honest proposals optimistically, \ie, when honest participation is high. Moreover, due to the proposer selection mechanism satisfying uniqueness, it can do so without increasing the slot length to $4\Delta$ rounds, \emph{if the optimistic assumption is expanded to also assume that network latency is $\frac{\Delta}{2}$}. We discuss this at length in Appendix~\ref{appendix-fast-conf}.

Finally, observe that, if $\eta > 1$, enabling subsampling allows for ex-ante reorgs~\cite{3attacks}. Since reorg resilience is central in the security analysis of propose-vote-merge protocols, this leads to entirely different security arguments being necessary, and consequently to reduced adversarial tolerance.

\begin{algo}
\vbox{
\small
\begin{numbertabbing}\reset
  xxxx\=xxxx\=xxxx\=xxxx\=xxxx\=xxxx\=MMMMMMMMMMMMMMMMMMM\=\kill
  \textbf{function} $\ulmdghost(\V,t)$ \label{}\\
  \> \textbf{return} $\ghost(\FIL_{\text{rlmd}}(\V, t), t)$ \label{}\\
  \textbf{function} $\FIL_{\text{rlmd}}(\V, t)$ \label{}\\
  \> \textbf{return} $\FIL_{\text{lmd}}(\FIL_{\eta\text{-exp}}(\FIL_{eq}(\V,t)))$ \label{}\\[-5ex]
\end{numbertabbing}
}
\caption{$\ulmdghost$ Fork-Choice function}
\end{algo}

\subsection{Tradeoff between dynamic availability and resilience to asynchrony}
The expiry parameter $\eta$ allows us to explore a tradeoff space between dynamic availability and resilience to asynchrony. At the extremes, $\eta = \infty$ gives us \LMDGHOST, a protocol which achieves asynchronous safety, but requires $> \frac{n}{2}$ honest participants to always be awake, and $\eta = 1$ gives us \Goldfish, a dynamically available protocol which does not tolerate even a single slot of asynchrony. As we show in Section~\ref{sec:ulmd-properties}, \ULMDGHOST with $1 < \eta < \infty$ sits somewhere in between, achieving weaker forms of both properties, namely $\eta$-dynamic-availability and $(\eta, \eta -1)$-asynchrony-resilience, \ie, it can tolerate $< \eta - 1$ slots of asynchrony, but it is only secure with the stronger assumptions of the $\eta$-sleepy model. The greater resilience to asynchrony is unsurprisingly due to the longer expiration period, which allows the latest messages of honest validators to persist even if periods of asynchrony prevent those validators from renewing their votes in the views of other validators. The same considerations apply to sudden drops in active participation, violating $\eta$-sleepiness assumptions. Even temporary violations of the basic $1$-sleepiness assumption, \ie, a \emph{temporary adversarial majority} (for example if all honest validators are asleep), can be tolerated. In all such cases, the longer expiry period prevents the set of active validators whose votes are unexpired, and thus relevant to the fork-choice function, from suddenly shrinking or disappearing altogether. On the other hand, taking into account votes of validators which might be asleep weakens dynamic availability, as we have seen for \LMDGHOST in Theorem~\ref{thm:lmd-not-da}, motivating why the $\eta$-sleepiness assumption is required.

\subsection{Properties}
\label{sec:ulmd-properties}

We start by showing that Proposition~\ref{prop:induction-step-assumption} holds in $\eta$-compliant executions of \ULMDGHOST. 

\begin{lemma}
\label{thm:ulmd-induction-prop-applies}
Proposition~\ref{prop:induction-step-assumption} holds for \ULMDGHOST in $\eta$-compliant executions.
\end{lemma}

\begin{proof}
Let $\V$ be the view of an active validator at a round $\in \{3\Delta t, 3\Delta t + \Delta\}$. By the synchrony assumption, and since the buffer is merged at round $3\Delta (t-1) + 2\Delta$, all honest votes from slot $t-1$ are in $\V$ and, by assumption, they are for descendants of $B$. The only votes to consider in order to decide whether $B$ is canonical in $\V$ are those from slots $\in [t-\eta, t-1]$, because votes from slots prior to $t - \eta$ are expired at slot $t$. Votes that are not for descendants of $B$ might be those from adversarial validators in $A_{t}$, or from validators in $H_{t- \eta, t-2} \setminus H_{t-1}$, \ie, those that have voted in at least some slot $\in [t-\eta, t-2]$, but did not vote in slot $s-1$. Observe that $H_{t-1} \cap A_t$ might not be empty; there might be validators that were active in slot $t-1$ but were (shortly after) corrupted. Therefore, $\V$ might contain more than one vote from slot $t-1$ from some of these validators. 

Let $E \subset H_{t-1} \cap A_t$ be the set of validators in $H_{t-1} \cap A_t$ for which $\V$ contains more than one vote from slot $t-1$. Due to equivocation discounting, votes from validators in $E$ will not count. Observe that the number of votes that are not for descendants of $B$ and that are counted in $\V$ is upper bounded by $|(A_t\setminus E) \cup (H_{t-\eta, t-2} \setminus H_{t-1})| = |(A_t \cup (H_{t-\eta, t-2} \setminus H_{t-1})) \setminus E| = |A_t \cup (H_{t-\eta, t-2} \setminus H_{t-1})| - |E|$, where the first equality follows from $E \subset H_{t-1}$. Since $\V$ contains votes for descendants of $B$ for all validators in $H_{t-1}$, the number of votes for descendants of $B$ and that are counted in $\V$ is lower bounded by $|H_{t-1} \setminus E| = |H_{t-1}| - |E|$. Since this is an $\eta$-compliant execution, $\eta$-sleepiness holds, \ie, $|H_{t-1}| > |A_t \cup (H_{t-\eta, t-2} \setminus H_{t-1}|)$, so $B$ is canonical in $\V$. 
\end{proof}

As shown in Section~\ref{sec:pvm}, since Proposition~\ref{prop:induction-step-assumption} holds for \ULMDGHOST in $\eta$-compliant executions, the next two theorems follow. Observe that, by the hierarchy of sleepy models (see Section~\ref{sec:model}), the following results are also satisfied for $\tau \geq \eta$. In Appendix~\ref{appendix:limitations}, we show that these results are tight.

\begin{theorem}[Reorg resilience]
\label{thm:ulmd-reorg-resilience}
\ULMDGHOST is $\eta$-reorg-resilient.
\end{theorem}

\begin{theorem}[Dynamic availability]
\label{thm:ulmd-dynamic-availability}
\ULMDGHOST is $\eta$-dynamically-available.
\end{theorem}

\begin{theorem}[Asynchrony resilience]
\label{thm:ulmd-asynchrony-resilience}
\ULMDGHOST is $(\eta, \eta-1)$-asynchrony-resilient.
\end{theorem}

\begin{proof}
Consider an $(\eta, \eta-1)$-compliant execution, with a $\varTPA{(\eta-1)}$ $(t_1, t_2)$, and an honest proposal $B$ from a slot $t \leq t_1$ after $\GST + \Delta$. First, since synchrony holds for slots $[t,t_1]$, and thus network synchrony holds until round $3\Delta t_1 + 2\Delta$, all the properties of \ULMDGHOST hold until then, including reorg resilience. In particular, starting from round $\geq 3\Delta t + \Delta$, $B$ is in the canonical chain of all active validators in those slots, as they coincide with the aware validators. We then only need to consider aware validators at slots $s > t_1$.
Suppose $B$ is in the canonical chain of all aware validators at all slots $< s$. In particular, $B \preceq \chain_i^r$ for a validator $v_i \in H_{t_1}$ which is active at a round $r \in [3\Delta t_1 + \Delta, 3\Delta (s-1) + \Delta]$, because validators in $H_{t_1}$ are always aware when active. Therefore, validators in $H_{t_1} \setminus A_{s}$ only ever broadcast votes for descendants of $B$ in slots $[t_1, s-1]$.

Consider first the case $s \in (t_1, t_2]$. Then, the aware validators at a round $r \in \{3\Delta s, 3\Delta s + \Delta\}$ are exactly the validators $H_{t_1}$ which are active in $r$. Consider then such a validator $v_i \in H_{t_1}$, and its view $\V_i$ at round $r$. View $\V_i$ contains all honest votes from slot $t_1$ because, by definition of $\varTPA{(\eta-1)}$, $v_i$ was awake at round $3\Delta t_1 + 2\Delta$, at which point it received all honest votes from slot $t_1$ and merged them into its view. Such votes are not expired at slot $s$, since $t_2 - t_1 \leq \eta-1$ implies $t_1 > t_2-\eta \geq s - \eta$, \ie, $t_1$ is within the expiry period $[s -\eta, s-1]$ for slot $s$. All validators in $H_{t_1} \setminus A_s$ are not equivocators in $\V_i$, since they are not corrupted by round $3\Delta s + \Delta$. Therefore, their latest votes in $\V_i$ all count for a descendant of $B$ in $\V_i$. The other votes which are counted in $\V_i$ are those from $A_{s}$ and $H_{s-\eta, s-1}\setminus H_{t_1}$. Since the execution is $(\eta, \eta-1)$-compliant, we have $|H_{t_1}\setminus A_s| > |A_s \cup (H_{s-\eta, s-1}\setminus H_{t_1})|$, and thus $B$ is canonical in $\V_i$.

Consider now the case $s = t_2+1$. Now, aware views coincide with active views, so we let $\V_j$ be the view of an active validator $v_j$ at a round $r \in \{3\Delta (t_2+1), 3\Delta (t_2+1) + \Delta\}$. Since synchrony holds from slot $t_2$, view $\V_j$ contains all latest votes from $H_{t_1} \setminus A_{t_2+1}$, which are all from slots $[t_1, t_2]$, and thus for descendants of $B$ by assumption. Moreover, $t_1 \geq t_2 - (\eta - 1) = (t_2 + 1) - \eta$, so all such votes are not expired at slot $t_2 + 1$. We can again conclude that $B$ is canonical in $\V_j$, because $|H_{t_1}| > |A_{s} \cup (H_{s-\eta, s-1}\setminus H_{t_1})|$ holds for $s = t_2+1$ as well.

Finally, suppose $s > t_2+1$. Since aware and active views coincide at slot $s-1$, $B$ is canonical in all active views at slot $s-1$ by assumption, so all honest votes from slot $s-1$ are for a descendant of $B$. Since synchrony holds as well, we can apply~\ref{prop:induction-step-assumption} and conclude that $B$ is canonical at all active views at slot $s$. 
\end{proof}

For \ULMDGHOST with $\eta \leq 2$, Theorem~\ref{thm:ulmd-asynchrony-resilience} does not say anything, because a $\piTPA$ is empty for $\pi \leq 1$. For $\eta = 1$, this is entirely to be expected, given the limitations of \Goldfish in this sense.

\section{Related works}
\label{sec:related-works}

The sleepy model of consensus, introduced by Pass and Shi~\cite{sleepy}, abstracts the functioning implicitly introduced by the Bitcoin protocol~\cite{nakamoto2008bitcoin}, modeling a distributed system in which participants can be either online or offline; in other words, where the participation is \emph{dynamic}, assuming only that at any given time a majority of online participants are honest. Moreover, Pass and Shi show that the longest chain protocol\footnote{Inspired by the Bitcoin protocol, but replacing leader election through proof of work with a VRF-based election} is safe and live within its model, \ie, that it is dynamically available. Longest chain protocols are also meaningfully resilient to temporary asynchrony: the deeper a block is in the longest chain, the longer the period of asynchrony required for it to be reorged. In other words, blocks \emph{accumulate safety over time}. On the other end, longest chain protocols are not reorg resilient, as newly proposed blocks can displace those at the tip of the chain.

Other than longest chain, to our knowledge only two other types of dynamically available protocols have been designed so far. One is \Goldfish~\cite{goldfish}, a variant of \LMDGHOST~\cite{gasper, zamfir} which we have discussed extensively in this work. Unlike longest chain protocols, \Goldfish is reorg resilient, but not at all tolerant of asynchrony. Finally, another class of approaches to integrate dynamic participation within a consensus protocol has been recently devised~\cite{DBLP:conf/ccs/Momose022, DBLP:journals/iacr/MalkhiMR22}. Similarly to Goldfish, these do not tolerate even temporary asynchrony, raising the same questions about the practicality of their usage. 

% Buterin \emph{et al.}~\cite{gasper} devise Gasper, a proof-of-stake consensus protocol consisting of two protocols: Casper~\cite{casper}, a gadget on top of a block proposal mechanism whose role is to finalize blocks, and \LMDGHOST, a synchronous consensus protocol that allows for dynamic participation. 

% {Neu, Tas, and Tse~\cite{DBLP:conf/sp/NeuTT21} prove that, although Gasper can be expressed through (a variation of) the sleepy model, its synchronous component, \LMDGHOST, is actually not dynamically available. They do so by presenting an attack to its liveness.}

% D'Amato, Neu, Tas, and Tse~\cite{goldfish} later introduce \Goldfish, a simplified variant of \LMDGHOST, and prove that it is secure \emph{and} reorg resilient under similar assumptions to those of the sleepy model. As we have seen

Momose and Ren~\cite{DBLP:conf/ccs/Momose022} present a quorum-based atomic broadcast protocol in the sleepy model that simultaneously supports dynamic participation, albeit requiring periods of stable participation \emph{for liveness}, and achieves constant latency. They extend the classic BFT approach from static quorum size to \emph{dynamic quorum} size, \ie, according to the current participation level, while preserving properties of static quorum. In particular, their protocol is built upon a graded agreement protocol using a quorum-based design. 

Malkhi, Momose, and Ren~\cite{DBLP:journals/iacr/MalkhiMR22} improves on the latency of Momose and Ren~\cite{DBLP:conf/ccs/Momose022}, and present a synchronous protocol that is live under \emph{fully fluctuating participation}. They present a Byzantine atomic broadcast protocol in the sleepy model~\cite{sleepy} with 3 round latency, but tolerating only one-third Byzantine nodes, rather than one-half. This solution has been recently improved~\cite{min-corrup-fluctuating} achieving optimal $\frac{1}{2}$ corruption threshold.

% Beyond just dynamically available protocols in isolation, their combination with \emph{accountability gadget} (also known as finality gadgets) has been analyzed extensively. 

% Buterin \emph{et al.}~\cite{gasper} devise Gasper, a proof-of-stake consensus protocol consisting of two protocols: Casper~\cite{casper}, a gadget on top of a block proposal mechanism whose role is to finalize blocks, and \LMDGHOST, a synchronous consensus protocol. Motivated by the analysis of Gasper, Neu, Tas, and Tse~\cite{DBLP:conf/sp/NeuTT21} formulate the notion of \emph{ebb-and-flow protocol}, capturing the desired properties of such a composite protocol, and give a first secure construction for such a protocol. 

\section{Conclusion and future work}

\label{sec:conclusion}

Dynamically available protocols have recently been explored in the context of blockchain protocols, based on (variants of) the sleepy model~\cite{sleepy}. In this work we presented a generalization of this model, with more generalized and stronger constraints in the corruption and sleepiness power of the adversary, and we formally proved properties and limitations of (a variant of) \LMDGHOST~\cite{zamfir}, the dynamically available component of Gasper~\cite{gasper}, \Goldfish~\cite{goldfish}, a synchronous dynamically available and reorg resilient protocol, and \ULMDGHOST, our novel protocol. Table~\ref{table:sum} summarizes the properties achieved by these protocols. \ULMDGHOST results in a provably secure synchronous dynamically available protocol that can be regarded as a potential future substitute for the existing \LMDGHOST component in Gasper. 

%\ULMDGHOST results in a synchronous protocol that has interesting practical properties: it is $\eta$-dynamically available (Theorem~\ref{thm:ulmd-dynamic-availability}) and $\eta$-reorg resilient (Theorem~\ref{thm:ulmd-reorg-resilience})), with $\eta$ being the expiry period in which votes are considered to make protocol's decisions, and it is resilient to asynchronous periods lasting less than $\eta-1$ slots. Because of these properties, we believe that \ULMDGHOST can be considered a viable future replacement the current \LMDGHOST component of Gasper, achieving a sufficient degree of both dynamic availability and asynchrony resilience.

%\subsection{Future work}

{A characterization through components of Gasper has been first formalized by Neu, Tas, and Tse~\cite{DBLP:conf/sp/NeuTT21}. Motivated by understanding Gasper~\cite{gasper} design goals, the authors introduce the \emph{partially synchronous sleepy model}, and then define the desiderata of the Ethereum's consensus protocol through the notion of an \emph{ebb-and-flow} protocol. In the partially synchronous sleepy model, (i) before a global stabilization time ($\GST$) message delays are chosen by an adversary, and after that the network becomes synchronous, with delay upper-bound $\Delta$, and (ii) before a global awake time ($\GAT$) the adversary can set any sleeping schedule for the participants, and after that all honest participants become awake. In particular, a (secure) ebb-and-flow protocol is constituted by both a dynamically available protocol that, if $\GST = 0$, \ie, under synchrony, is guaranteed to be safe and live at all times (according to Definition~\ref{def:security}), \emph{and} a finalizing protocol that is guaranteed to be safe at all times, and live after $\max\{\GST, \GAT\}$. In the context of Gasper, the former is represented by \LMDGHOST, while the latter by Casper~\cite{casper}. Understanding the interaction between \ULMDGHOST and a finality component is an interesting open problem, especially with respect to asynchrony. Specifically, we reserve for future research the development of a secure ebb-and-flow protocol which incorporates \ULMDGHOST as a dynamically available component, along with a partially synchronous finality component. Moreover, it is an open question whether the techniques of this work can be applied to the family of quorum-based dynamically available protocols~\cite{DBLP:conf/ccs/Momose022, DBLP:journals/iacr/MalkhiMR22}, to strengthen their resilience of asynchrony at the cost of weakened tolerance to dynamic participation, as we have done with \Goldfish.}

\begin{table}[t!]
\centering
% Removed the \footnotesize for a bigger font size.
\begin{tabular}{|p{0.3\columnwidth}|p{0.3\columnwidth}|p{0.3\columnwidth}|}
\hline
 & \multicolumn{1}{c|}{\textbf{Dynamic Availability}} & \multicolumn{1}{c|}{\textbf{Asynchrony Resilience}} \\ \hline
\centering\raisebox{-0.8\height}{\LMDGHOST}  & \ding{55} (only $\tau = \infty$, Theorem~\ref{thm:lmd-not-da}) & \ding{51} ($\tau, \pi = \infty$, Theorem~\ref{thm:ulmd-asynchrony-resilience}) \\ [1ex] \hline % added [1ex] for vertical padding
\centering\raisebox{-0.8\height}{\Goldfish}  & \ding{51} ($\tau = 1$, Theorem~\ref{thm:ulmd-dynamic-availability}) & \ding{55} (Theorem~\ref{thm:goldfish-not-asynchrony-resilient}) \\ [1ex] \hline
\centering{\ULMDGHOST, expiry period $\eta$} & \ding{51} ($\tau = \eta$, Theorem~\ref{thm:ulmd-dynamic-availability}) & \ding{51} ($\tau, \pi = \eta, \eta -1$, Theorem~\ref{thm:ulmd-asynchrony-resilience}) \\ [1ex] \hline
\end{tabular}
\vspace{0.7ex}
\caption{\textnormal{The properties achieved by \LMDGHOST with view-merge and no subsampling (Section~\ref{sec:lmd-ghost}), \Goldfish (Section~\ref{sec:goldfish}), and \ULMDGHOST with expiry period $\eta \in (1, \infty)$ (Section~\ref{sec:ULMD}).}}
\label{table:sum}
\vspace{-2em}
\end{table}

\appendix
\bibliography{references}
\bibliographystyle{plainurl}% the mandatory bibstyle
\section{Limitations of \ULMDGHOST}
\label{appendix:limitations}

We first show limitations on the reorg resilience of \ULMDGHOST when $\eta > \tau$, except for $\eta = 1$, \ie, \Goldfish. We construct a $\tau$-compliant execution in which old, unexpired votes of honest validators which are currently asleep are used by the adversary to fuel a reorg. 

\begin{theorem}
\ULMDGHOST is not $\tau$-reorg-resilient for any $1 \leq \tau < \eta$.
\end{theorem}

\begin{proof}
We prove this theorem with an example. Let us consider a validator set of size $n = 2m+1$, partitioned in three sets, $V_1$, $V_2$, and $V_3$, with $V_1 = \{v_1\}$, $|V_2| = m+1$, $|V_3| = m-1$. Validators in $V_2$ and $V_3$ are all initially honest, while $v_1$ is adversarial. Let $t-1$ and $t$ be two adversarial slots, \ie, controlled by $v_1$. In slot $t$, validator $v_1$ publishes conflicting blocks $A$ and $B$, one as a proposal for slot $t-1$ and the other for slot $t$. By round $3\Delta t + \Delta$, the adversary delivers only $A$ to validators in $V_2$, and only $B$ to validators in $V_3$, so that the former vote for $A$ and the latter for $B$ in slot $t$. At this point, the adversary puts all validators in $V_3$ to sleep, and then does nothing until slot $t+\eta-1$. Meanwhile, validators in $V_2$ keep voting for $A$, since $V_2$ contains $m+1 > \frac{n}{2}$ validators, so $A$ stays canonical in all of the views of every member of $V_2$. Suppose in particular that the proposer of slot $t+1$ is in $V_2$, so that it makes a proposal $C$ extending $A$. We now show that the adversary can induce a reorg of $C$, exploiting the votes of the asleep validators $V_3$, so that reorg resilience is not satisfied in this execution. We then only have to show that the execution is $\tau$-compliant, in order to show that the protocol is not $\tau$-reorg-resilient.

At the voting round $3\Delta (t+\eta-1) + \Delta$, the adversary votes for $B$ with $v_1$. After the voting round, it corrupts two validators $v_2, v_3 \in V_2$, and starts voting for $B$ with them, broadcasting late votes for slot $t + \eta - 1$. These votes are delivered to all awake validators by round $3\Delta (t+\eta-1) + 2\Delta$, and are therefore in all of their views at the voting round of slot $t+\eta$. The votes of $v_2$ and $v_3$ are equivocations, so they are discounted, both for $B$ and for $A$. Slot $t$ is in $[t, t + \eta)$, the expiration period for slot $t + \eta$, so the votes of $V_3$ count at this slot. Therefore, in all views of the remaining honest validators in $V_2$, $B$ has $m$ votes, \ie, those of $V_3$ and $v_1$, and descendants of $A$ only $m-1$, because two have been discounted. $B$ is then canonical in such views, and reorg resilience (of $C$) is violated. The adversary does not perform any more corruptions nor puts to sleep any more validators, and does not wake up validators in $V_3$.

We now show that this execution is $\tau$-compliant. For any slot $s$, we show that $\tau$-sleepiness holds at slot $s$, \ie, that Equation~\ref{eq:sleepy-req} holds. For any slot $s \leq t+1$, this is clear, because we have $|H_{s-1}| = |V_2 \cup V_3| = 2m > 1 = |V_1| = |A_s \cup (H_{s-\tau, s-2}\setminus H_{s-1})|$. For any slot $s \in [t+2, t+\eta -1]$, we have $H_{s-1} = V_2$ and $A_s = V_1$, because the two corruptions only happen after round $3\Delta (t+\eta-1) + \Delta$. Therefore, $|H_{s-1}| = |V_2| = m+1 > m = |V_1 \cup V_3| \geq |A_s \cup (H_{s-\tau, s-2}\setminus H_{s-1})|$, so $\tau$-sleepiness is satisfied. 
For any slot $s \geq t+\eta$, we have $A_s = \{v_1, v_2, v_3\}$, $V_2 \setminus \{v_2, v_3\} \subseteq H_{s-1}$ and $H_{s-\tau, s-2}\setminus H_{s-1} = \emptyset$, because $\eta > \tau$ implies $s-\tau \geq t + \eta - \tau > t$, so $V_3 \cap H_{s-\tau, s-2} = \emptyset$. $\tau$-sleepiness is then satisfied, because $|H_{s-1}| \geq m-1 > 3 = |A_s| =  |A_{s} \cup H_{s-\tau, s-2}\setminus H_{s-1}|$. Since the execution is $\tau$-compliant, but does not satisfy reorg resilience, the protocol is not $\tau$-reorg-resilient.
\end{proof}

The second limitative result we present concerns dynamic availability. Unsurprisingly, an expiry period $\eta > \tau$ means that \ULMDGHOST is also not $\tau$-dynamically-available. 

\begin{theorem}
\ULMDGHOST is not $\tau$-dynamically-available for any $1 \leq \tau < \eta$ and for any confirmation rule with $\Tconf < \lfloor \frac{n-5}{4}\rfloor\eta = O(\eta\cdot n)$ slots. In particular, it is not $\tau$-dynamically available with the $\kappa$-deep confirmation rule, for $\kappa < \lfloor\frac{n-5}{4}\rfloor\eta$.
\end{theorem}

\begin{proof}
Consider a validator set of size $n = 2m+1$, partitioned in three sets, $\mathcal{C}_0$, $\mathcal{A}_0$, and $\mathcal{S}_0$, standing for \emph{corrupted, active, and sleepy}, respectively, with $\mathcal{C}_0 = \{v_1\}$, $|\mathcal{A}_0| = m+2$, $|\mathcal{S}_0| = m-2$. Validators in $\mathcal{A}_0$ and $\mathcal{S}_0$ are all initially honest, while $v_1$ is adversarial. Let $t-1$ and $t$ be two adversarial slots, \ie, controlled by $v_1$. In slot $t$, validator $v_1$ publishes conflicting blocks $A$ and $B$, one as a proposal for slot $t-1$ and the other for slot $t$. By round $3\Delta t + \Delta$, the adversary delivers only $A$ to validators in $\mathcal{A}_0$, and only $B$ to validators in $\mathcal{S}_0$, so that the former vote for $A$ and the latter for $B$ in slot~$t$. 

At this point, the adversary puts all validators in $\mathcal{S}_0$ to sleep, and then does nothing until immediately after round $3\Delta (t+\eta -2) + \Delta$, \ie, the voting round of slot $t+\eta-2$, at which point it corrupts two validators $\{v_2, v_3\} \in \mathcal{A}_0$. Up until this point, all validators in $\mathcal{A}_0$ have kept voting for $A$, since $|\mathcal{A}_0| = m+2 > \frac{n}{2}$ validators. At slot $t+\eta -1$, the adversarial validators initially do not broadcast votes. By round $3\Delta (t+\eta-1) + 2\Delta$, the adversary wakes up validators in $\mathcal{S}_0$, so that they are active in slot $t+\eta$. At slot $t+\eta$, the adversary publishes three votes for $B$ from slot $t+\eta -1$, from validators $\{v_1, v_2, v_3\}$. By round $3\Delta (t+\eta) + \Delta$, it delivers these only to validators in $\mathcal{A}_0$. Since the expiration period $[t, t + \eta - 1]$ for slot $t + \eta$ contains $t$, the votes of $\mathcal{S}_0$ count at slot $t + \eta$. Therefore, in the views of the validators in $\mathcal{S}_0$ at slot $t+\eta$, descendants of $A$ have a total of $m+2$ votes, from all validators in $\mathcal{A}_0$, including the newly corrupted ones, while $B$ only has $m-2$ votes from $\mathcal{S}_0$. Thus, $A$ is canonical in their views, and they vote for it. The views of the $m$ remaining honest validators in $\mathcal{A}_0$ also include the three adversarial votes for $B$ from slot $t+\eta -1$, so descendants of $A$ only have $m$ votes, while $B$ has $m+1$. $B$ then is canonical in their views, and they vote for it. The three adversarial validators also do so, so $B$ receives $m+3$ votes and is canonical in the following slots. After the voting round of slot $t+\eta$, the adversary then puts all but two of the $m-2$ validators in $\mathcal{S}_0$ to sleep.

In slot $t + \eta + 1$, there are then $m+2$ active validators: the two which are still active from $\mathcal{S}_0$, and $m$ which are still honest from $\mathcal{A}_0$. There are also three adversarial validators and $m-4$ validators from $\mathcal{S}_0$ asleep from the previous slot. We are therefore in the same situation as in slot $t+1$, except we have two more adversarial validators (from $\mathcal{A}_0$) and two less asleep validators (from $\mathcal{S}_0$). We let $\mathcal{C}_1$ be the three adversarial validators, $\mathcal{A}_1$ be the $m+2$ active validators and $\mathcal{S}_1$ be the $m-4$ asleep validators.
The adversary repeats the same pattern. It corrupts two more validators from $\mathcal{A}_1$ after the voting round of slot $(t + \eta) + (\eta -2) = t + 2\eta - 2$, and at round $3\Delta (t+2\eta - 1) + 2\Delta$ wakes up all validators in $\mathcal{S}_0$ so that they are active by slot $t+2\eta$. It then votes with all of the five adversarial validators for the branch of $A$ at slot $t + 2\eta - 1$, but delivers such votes only to validators in $\mathcal{A}_1$ by the voting round of slot $t+2\eta$. Then, at slot $t + 2\eta$, the branch of $A$ has $m+1$ votes in the views of validators in $\mathcal{A}_1$, \ie, the adversarial votes plus the votes from the $m-4$ validators in $\mathcal{S}_1$, which were put to sleep after voting for $A$ at slot $t+\eta$. Therefore, it is canonical in their views and they vote for it, and so does the adversary. On the other hand, the views of validators in $\mathcal{S}_1$ at that round do not include the adversarial votes for $A$, and so all validators in $\mathcal{S}_1$ vote for~$B$.

All but \emph{four} of them are now put to sleep, so that at slot $t + 2\eta + 1$ there are $m+2$ active validators, $m-6$ asleep validators and five adversarial validators. We let these new sets of validators be $\mathcal{A}_2, \mathcal{S}_2, \mathcal{C}_2$, respectively. Again, the adversary has reorged from one branch to the other, while only needing to corrupt two asleep validators into two adversarial validators, and while otherwise preserving the same setup. They can repeat this until the number of adversarial validators reaches $m-1$, which does not allow for two additional corruptions. After the $k^{th}$ reorg, at slot $t+k\eta+1$, there are $m+2$ active validators $\mathcal{A}_k$, $2k+1$ adversarial validators $\mathcal{C}_k$, and $m - 2(k+1)$ asleep validators $\mathcal{S}_k$. Therefore, the adversary can repeat this up to $k \leq \lfloor \frac{m-2}{2} \rfloor = \lfloor \frac{n-5}{4}\rfloor$ times. Each time they do so, they can reorg from one branch to the other after $\eta$ slots, for a total of $\lfloor \frac{n-5}{4}\rfloor\eta$ slots. By assumption, $\Tconf < \lfloor \frac{n-5}{4}\rfloor\eta$ slots. If no confirmation has been made after $\Tconf$ slots, then liveness is violated. If one has been made, then the confirmed block can still be reorged by slot $\lfloor \frac{n-5}{4}\rfloor\eta$, and the conflicting branch eventually confirmed afterwards, violating safety.

To complete the proof, we only need to verify that $\tau$-sleepiness is satisfied. For slots $s \leq t$, we have $|H_{s-1}| = 2m > 1 = |A_{s} \cup (H_{s-\tau, s-2}\setminus H_{s-1})|$, so it is indeed satisfied. Consider now some $1 \leq k \leq \frac{m-2}{2}$, and slots $[t+(k-1)\eta + 1, t+k\eta]$.  For $s \in [t+(k-1)\eta + 1, t+k\eta-1]$, we have $|H_{s-1}| \geq |\mathcal{A}_k| = m+2$, $|A_s| \leq |\mathcal{C}_k| = 2k+1$ and $|H_{s-\tau, s-2}\setminus H_{s-1}| = |\mathcal{S}_{k-1}| = m-2k$, so $|H_{s-1}| \geq m+2 > m+1 = (2k+1) + (m - 2k) \geq |A_{s} \cup (H_{s-\tau, s-2}\setminus H_{s-1})|$, and $\tau$-sleepiness at slot $s$ is satisfied. For $s = t+k\eta$, we have $|H_{s-1}| = m$, because two more validators have been corrupted, $|A_s| = |\mathcal{C}_k| = 2k+1$, and $H_{s-\tau, s-2}\setminus H_{s-1} = \emptyset$, because $\tau < \eta$ implies $s - \tau = t+k\eta - \tau > t+(k-1)\eta$, which is the last slot in which $\mathcal{S}_{k-1}$ were active
. Since $2k+2 \leq m$, we have that $2k+1 < m$, so $\tau$-sleepiness at slot $t+ k\eta$ is indeed satisfied. In slots after the last reorg, all honest validators are active, and there are $\geq m + 2 > \frac{n}{2}$ of them, so $\tau$-sleepiness is also satisfied.
\end{proof}

\begin{theorem}
\label{thm:rlmd-not-asyn-res}
\ULMDGHOST with finite $\eta$ is not $(\tau, \pi)$-asynchrony-resilient for any $\tau > \pi \geq \max(\eta, 2)$, nor for $\tau = \pi = \infty$. 
\end{theorem}

\begin{proof}
We have to show that \ULMDGHOST is not asynchrony resilient for any $\tau > \pi \geq \eta$, which we do by showing that \ULMDGHOST is not $(\infty, \pi)$-asynchrony-resilient, by constructing an $(\infty, \eta)$-compliant execution in which asynchrony-resilience does not hold. Since $E_{\tau, \pi}$ is monotonically decreasing in $\tau$ and monotonically increasing in $\pi$, $E_{\infty, \eta} \subset E_{\tau, \pi}$ for any $\tau > \pi \geq \eta$, and similarly $E_{\infty, \eta} \subset E_{\infty, \infty}$, so the desired result follows. We consider a validator set $\{v_1, v_2, v_3\}$, where all validators are honest at all times, and consider an execution with a $\varTPA{\eta}$ $(t, t+\eta)$, which is $\neq \emptyset$ since $\eta \geq 2$. In particular, network synchrony does not hold at slot $t+\eta-1$. Before round $3\Delta (t+\eta - 1) + 2\Delta$, validator $v_3$ is asleep. It wakes up at that round, and stays awake thereafter, so $v_3 \in H_{s}$ for $s \geq t+\eta$. Both validators $v_1$ and $v_2$ are active at all rounds $\leq 3\Delta t + 2\Delta$, so $H_s = \{v_1, v_2\}$ for $s \leq t$. Validator~$v_1$ subsequently falls asleep, and only wakes up again in round $3\Delta (t+\eta) + 2\Delta$, while validator $v_2$ is always awake. Upon waking up at round $3\Delta (t+\eta -1) + 2\Delta$, validator $v_3$ does not see any message before merging the buffer into its view, due to asynchrony. Validator~$v_3$ is the proposer of slot $t+\eta$, and, due to the lack of messages in its view, it proposes a block $B$ extending $B_{\text{genesis}}$,  which conflicts with all previous honest proposals. Validator~$v_3$ then also votes for $B$ at slot $t+\eta$, while $v_2$ does not. All three honest validators are active at round $3\Delta (t+\eta) + 2\Delta$, so they receive these votes and merge them into their view. The latest vote from $v_1$ is from slot $t$, and is expired at slot $t+\eta+1$. Therefore, the only unexpired latest votes at the voting round of slot $t+\eta+1$ are those from $v_2$ and $v_3$ from slot $t+\eta$. If $B$ wins the tiebreaker, it is then canonical in the views of the three validators. All honest proposals from slots $\leq t$ are then not canonical in these active views, which are also aware views since we are at a slot $> t+\eta$, so asynchrony-resilience is not satisfied in this execution. In order to show the desired result, we then only need to show that the execution is $(\infty, \eta)$-compliant. For slots $s \not \in (t, t+\eta]$, we have to show that $\infty$-sleepiness holds. It suffices to show that $|H_{s-1}| \geq 2 > \frac{n}{2}$. For $s\leq t$, we have $H_{s-1} = \{v_1, v_2\}$, while for $s > t+\eta$ we have $\{v_2, v_3\} \subseteq H_{s-1}$ , so this is indeed the case. For slots $s \in (t, t+\eta+1]$, we have $H_{t} \setminus A_{s} = H_t = \{v_1, v_2\}$, so the condition which needs to hold during the $\varTPA{\eta}$ is satisfied. Moreover, $H_t$ are awake at round $3\Delta t + 2\Delta$, satisfying even the last condition of $(\infty, \eta)$-compliance. 
\end{proof}

\section{Fast confirmations}
\label{appendix-fast-conf}
We specify the protocol with fast confirmations and then analyze its properties. It requires a small modification to the generic propose-vote-merge protocol, changing the vote behavior so that validators vote \emph{as soon as they see a proposal, or at round $3\Delta t + \Delta$, whichever comes first} (which is also exactly the attestation behavior specified by the Ethereum consensus protocol). In the following, $\FC = \ulmdghost$.

\subsection{Protocol with fast confirmation}
In the following, and in Algorithm~\ref{alg:fast-conf}, we update the confirmed chain explicitly, contrary to Algorithm~\ref{alg:pvm-generic}. This is because the confirmed chain $\chainconf_i$ in the latter is at any point simply a function of the canonical chain $\chain_i$, \ie, $\chainconf_i = \chain_i^{\lceil \kappa}$. With fast confirmations, this is no longer the case. Moreover, in Algorithm~\ref{alg:fast-conf} we use $\FIL(\V, t).\V$ to refer to the view output by a filter. Observe that validators (which have synchronized clocks) update the variables $t$ and $r$ representing slot and round, respectively, through the protocol's execution.

\textsc{Propose}: Unchanged from Section~\ref{sec:pvm-protocol}

\textsc{Vote}: In rounds $[3\Delta t, 3\Delta t + \Delta]$, a validator $v_i$, upon receiving a proposal message [\textsc{propose}, $B$, $\V$, $t$, $v_p$] from $v_p$, merges its view with the proposed view $\V$. After doing so, or at round $3\Delta t + \Delta$ if no proposal is received, it updates its canonical chain by setting $\chain_i \gets \FC(\V_i, t)$, and broadcasts the vote message [\textsc{vote}, $\FC(\V_i, t)$, $t$, $v_i$]. 

\textsc{Confirm}: At round $3\Delta t + \Delta$, a validator $v_i$ merges its view with its buffer, \ie, $\V_i \gets \V_i \cup \B_i$, and sets $\B_i \gets \emptyset$. It then selects for fast confirmation the highest \emph{canonical} block $B_{\text{fast}} \prec \chain_i$  such that $\B_i$ contains $\geq \frac{2}{3}n$ votes from slot~$t$ for descendants of $B_{\text{fast}}$, from distinct validators. It then updates its confirmed chain $\mathsf{Ch}_i$ to the highest of $B_{\text{fast}}$ and $\chain_i^{\lceil \kappa}$, the $\kappa$-deep prefix of its canonical chain, \emph{as long as this does not result in updating $\mathsf{Ch}_i$ to some prefix of it} (we do not needlessly revert confirmations).

\textsc{Merge}: At round $3\Delta t + 2\Delta$, every validator $v_i$ merges its view with its buffer, \ie, $\V_i \gets \V_i \cup \B_i$, and sets $\B_i \gets \emptyset$.

\begin{algo*}[hbt!]
\vbox{
\small
\begin{numbertabbing}\reset

xxxx\=xxxx\=xxxx\=xxxx\=xxxx\=xxxx\=MMMMMMMMMMMMMMMMMMM\=\kill
    \textbf{State} \label{}\\
    \> \(\V_i \gets \{B_{\text{genesis}}\} \): view of validator $v_i$ \label{}\\
    \> \(\B_i \gets \emptyset \): buffer of validator $v_i$  \label{}\\
    \> \(\chain_i \gets B_{\text{genesis}}\): canonical chain of validator $v_i$ \label{}\\
    \> \(\chainconf_i \gets B_{\text{genesis}}\): confirmed chain of validator $v_i$ \label{}\\ 
    \> \(t \gets 0\): the current slot \label{}\\
    \> \(r \gets 0\): the current round \label{}\\
        \> $\textit{sentvote} \gets \textsc{false}$: indicates whether $v_i$ has voted in slot $t$ \label{}\\
    \textsc{propose}\\
    \textbf{at} $r=3\Delta t$ \textbf{do} \label{} \\
    \> \textbf{if} $v_i = v_p^t$ \textbf{then} \label{}\\
    \>\>$\V_i \gets \V_i \cup \B_i$ \label{}  \\
    \>\> $\B_i \gets \emptyset$ \label{}\\
    \>\> $B' \gets \FC(\V_i, t)$ \label{}\\
    \>\> $B \gets \mathsf{NewBlock}(B')$ 
    \` // append a new block on top of $B'$ \label{} \\
    \>\> $\chain_i \gets B$ \label{}\\
    \>\> send message [\textsc{propose}, $B$, $\V_i\,\cup \{B\}$, $t$, $v_i$] through gossip \label{} \\
     \textsc{vote and confirm}\\
     \textbf{at} $r=3\Delta t + \Delta$ \textbf{do} \label{}\\
     \> \textbf{if} $\neg \textit{sentvote}$ \textbf{then} \label{}\\
         \>\> $\chain_i \gets \FC(\V_i, t)$ \label{}\\
         \>\> send message [\textsc{vote}, $\FC(\V_i, t)$, $t$, $v_i$] through gossip \label{} \\
      \>\> $\textit{sentvote} \gets \textsc{true}$ \label{}\\
     \> $\V_i \gets \V_i \cup \B_i$ \label{}  \\
     \> $\B_i \gets \emptyset$ \label{}\\
     \> $B_{\text{fast}} \gets B_{\text{genesis}}$ \label{} \\
     \> $S_{\text{fast}} \gets \{B \prec \chain_i\colon |\{v_i\colon \exists B' \succ B  \ (\textsc{vote}, B', t, v_i) \in \V_i \}| \geq \frac{2}{3}n\}$ \label{} \\
     \> \textbf{if} $S_{\text{fast}} \neq \emptyset$ \textbf{then}: \label{}\\
    \>\> $B_{\text{fast}} \gets \underset{S_{\text{fast}}}{\text{arg max}} |B|$ \label{}\\
    \> \textbf{if} $\neg (B_{\text{fast}} \prec  \mathsf{Ch}_i \land \chain_i^{\lceil \kappa} \prec \mathsf{ch}_i) $ \textbf{then}: \label{}\\
     \>\> $\mathsf{Ch}_i \gets \underset{\chain \in \{\chain_i^{\lceil \kappa}, B_{\text{fast}}\}}{\text{arg max}} |\chain|$ \label{}\\
    \textsc{merge}\\
     \textbf{at} $r=3\Delta t + 2\Delta$ \textbf{do} \label{}\\
     \> $\V_i \gets \V_i \cup \B_i$ \label{}\\
       \> $\B_i \gets \emptyset$ \label{}\\
    \> $\textit{sentvote} \gets \textsc{false}$ \label{}\\
    \textbf{upon} receiving a gossiped message [\textsc{propose}, $B$, $\V$, $t$, $v_p^t$] \textbf{do} \label{}\\
    \> $\B_i \gets \B_i \cup \{B\}$ \label{}\\  
     \> \textbf{if} $\neg \textit{sentvote}$ \textbf{and} $r \in [3\Delta t, 3\Delta t + \Delta]$ \textbf{then} \label{}\\
     \>\>$\V_i \gets \V_i \cup \V$ \label{}\\
     \>\>$\chain_i \gets \FC(\V_i, t)$ \label{}\\
     \>\>send message [\textsc{vote}, $\FC(\V_i, t)$, $t$, $v_i$] through gossip \label{} \\
      \>\> $\textit{sentvote} \gets \textsc{true}$ \label{}\\
    \textbf{upon} receiving a gossiped message $V = $ [\textsc{vote}, $B$, $t'$, $v_i$] from \(v_i\) \textbf{do} \label{}\\
    \> $\B_i \gets \B_i \cup \{V\}$ \label{}\\  
    \textbf{upon} receiving a gossiped message $B = $ [\textsc{block}, $b$, $t'$, $v_i$] from \(v_i\) \textbf{do} \label{}\\
    \> $\B_i \gets \B_i \cup \{B\}$ \label{}\\[-5ex]
\end{numbertabbing}
}
\caption{Propose-vote-merge protocol for validator $v_i$}
\label{alg:fast-conf}
\end{algo*}

\subsection{Safety and liveness tradeoff}
The protocol with fast confirmations is safe when \emph{both} fast confirmations and standard ($\kappa$-deep) confirmations are safe, and live whenever \emph{at least one} is live. In particular, liveness is guaranteed by the liveness of the standard confirmations. On the other hand, the safety resilience of this protocol can be worse than $\frac{n}{2}$, which is what the original protocol tolerates \emph{when all honest validators are awake}, since $\eta$-sleepiness reduces to $|H_{t-1}| > |A_t|$. In the previous section we have in particular specified fast confirmations to require a quorum of size $\frac{2}{3}n$, which results in both liveness and safety resilience \emph{of fast confirmations} of $\frac{n}{3}$. We have chosen this quorum size because we are interested in using \ULMDGHOST in combination with a finality gadget, following the pattern of~\cite{DBLP:conf/sp/NeuTT21, aadilemma}, in which confirmations of the available protocol are input to the gadget, to preserve dynamic availability of the combined protocol. In this setting, we then want fast confirmations to require no further assumptions compared to the finality gadget, so that they can be live whenever the conditions are right for the gadget to be live, speeding up its action~\footnote{In this setting, we would then also consider doing away with the extra optimistic assumption about network latency, and going back to $4\Delta$ rounds instead, so that fast confirmations are always live after $\max(\GST, \GAT)$, without needing network latency $\leq \frac{\Delta}{2}$.}.
Due to this choice, safety resilience of the resulting protocol is then reduced to $\frac{n}{3}$ as well (\cf \Goldfish, where the chosen quorum is $\frac{3}{4}n$, so that the final protocol is still safe for $f < \frac{n}{2}$). On the other hand, we show in the next section that violating safety of a fast confirmation with quorum $\frac{2}{3}n$ requires $\frac{n}{3}$ equivocations, thus also making $\frac{n}{3}$ validators slashable. Therefore, the security guarantee of the resulting protocol \emph{under network synchrony} is that a safety violation requires \emph{either} safety of standard confirmations to be violated, implying a violation of $\eta$-sleepiness, and in particular $f \geq \frac{n}{2}$ if all honest validators are awake, \emph{or} $\frac{n}{3}$ adversarial validators have to be slashable for equivocation. All safety results in the next section follow this formulation.

\subsection{Properties}
Other than the small modifications we have made, the protocol with fast confirmation behaves exactly as the original protocol, because the confirmation rule does not in any way influence the protocol execution. Therefore, properties like reorg resilience and asynchrony resilience are preserved (up to accounting for those changes in the proofs). In the following, we then only discuss results for which the confirmation rule is relevant, \ie security results. Firstly, we show a result that fulfills the same role of the view-merge property (Lemma~\ref{thm:view-merge-property}) in our security analysis of fast confirmations, in the sense that it provides the base case for the induction of Theorem~\ref{thm:pvm-reorg-resilience}, allowing us to prove an analogous reorg resilience result for fast confirmations, which implies safety. Since liveness is obtained for free from the liveness of the protocol without fast confirmations, this shows $\eta$-dynamic-availability. Finally, we show that fast confirmations are themselves live when there are at least $\frac{2}{3}n$ honest validators awake and the real network latency is $\leq \frac{\Delta}{2}$. To make it easier to state the results, we work here with a slightly modified definition of $\tau$-compliant execution. In addition to satisfying $\tau$-sleepiness, we require that in no honest view $\geq \frac{n}{3}$ validators are seen as equivocators. This is a very weak requirement, since equivocation is a slashable offense.

\begin{lemma}
\label{thm:fast-base-case} Suppose network synchrony holds for rounds $[3\Delta t + \Delta, 3\Delta t + 2\Delta]$, and that an honest validator fast confirms block $B$ at slot $t$. Suppose also that, in the view of any active validator at slot $t+1$, $< \frac{n}{3}$ validators are seen as equivocators.  Then, all honest voters of slot $t+1$ vote for descendants of $B$.
\end{lemma}
\begin{proof}
Upon fast confirming $B$ at round $3\Delta t + \Delta$, the honest validator broadcasts $B$ and all votes $\geq \frac{2}{3}n$ votes which are responsible for the fast confirmation, so that they are in the view of all awake validators at round $3\Delta t + 2\Delta$, by synchrony. Therefore, they are also in the view of all active validators at round $3\Delta (t+1) + \Delta$. Consider one such view $\V$. By assumption, $< \frac{n}{3}$ validators are seen as equivocators in $\V$, so over $\frac{n}{3}$ out of the $\frac{2}{3}n$ votes are not discounted. Since they are from slot $t$, they are latest votes, and are the ones which count for the respective validators. Therefore, $w(B, \FIL_{ulmd}(\V, t+1).{\V}) > \frac{n}{3}$. On the other hand, $\V$ contains at most $\frac{n}{3}$ votes from slot $t$, conflicting with $B$ and by a validator which is not seen as an equivocator in $\V$. Therefore, $w(B', \FIL_{ulmd}(\V, t+1).{\V}) \leq \frac{n}{3}$ for any $B'$ conflicting with $B$, so $B$ is canonical in~$\V$, and an active validator with view $\V$ votes for a descendant of $B$.
\end{proof}

\begin{theorem}[Reorg resilience of fast confirmations]
\label{thm:fast-confirmation-persistence}
Consider an $\eta$-compliant execution of \ULMDGHOST. A block fast confirmed by an honest validator at a slot $t$ after $\GST$ is always in the canonical chain of all active validators at rounds $\geq 3\Delta (t+1) + \Delta$.
\end{theorem}

\begin{proof}
The proof follows that of Theorem~\ref{thm:pvm-reorg-resilience}, using Lemma~\ref{thm:fast-base-case} instead of Lemma~\ref{thm:view-merge-property} as the base case. The assumption of Lemma~\ref{thm:fast-base-case} about equivocators is satisfied by (the new) definition of $\eta$-compliance. Proposition~\ref{prop:induction-step-assumption}, which we have proven for $\eta$-compliant executions of \ULMDGHOST in Lemma~\ref{thm:ulmd-induction-prop-applies}, is still used for the inductive step. 
\end{proof}

\begin{theorem}[Dynamic availability]
\ULMDGHOST with fast confirmations is $\eta$-dynamically-available.
\end{theorem}

\begin{proof}
$\eta$-liveness follows directly from Theorem~\ref{thm:ulmd-dynamic-availability}, in particular from $\eta$-liveness of \ULMDGHOST without fast confirmations. This is because fast confirmations are not needed for the confirmed chain to make progress, and so liveness of the standard confirmations suffices. We then only need to show that it satisfies $\eta$-safety. If an honest validator fast confirms a block $B$ at slot $t$ in an $\eta$-compliant execution, then $B$ is in the canonical chain of all active validators at rounds $\geq 3\Delta(t+1) + \Delta$, by Theorem~\ref{thm:fast-confirmation-persistence}.
At slot $t+\kappa$, $B$ is then in the $\kappa$-slots-deep prefix of the canonical chain of all active validators and thus confirmed by them with the standard confirmation rule. Therefore, a safety violation involving conflicting confirmed chains $\chainconf_i^r$
and $\chainconf_j^{r'}$ can be reduced to a safety violation for the standard confirmation rule, for rounds $r+3\Delta(\kappa+1)$ and $r'+3\Delta (\kappa+1)$. Theorem~\ref{thm:ulmd-dynamic-availability} then implies the $\eta$-safety of the protocol.
\end{proof}

\begin{theorem}[Liveness of fast confirmations]
An honest proposal $B$ from a slot $t$ after $\GST + \Delta$ in which $|H_t| \geq \frac{2}{3}n$ and network latency is $\leq \frac{\Delta}{2}$ is fast confirmed by all active validators at round $3\Delta t + \Delta$.
\end{theorem}

\begin{proof}
Firstly, note that validators in $H_t$ are active in all rounds $[3\Delta (t_1) + 2\Delta, 3\Delta t + \Delta]$, because falling asleep at any point in those rounds would force them to go through the joining protocol again, and thus they would not be active prior to at least round $3\Delta t + 2\Delta$. Since network latency is $\leq \frac{\Delta}{2}$, all validators in $H_t$ receive the honest proposal by round $3\Delta t + \frac{\Delta}{2}$. By the view-merge property, Lemma~\ref{thm:view-merge-property}, they all vote for $B$. Again by the assumption on network latency, they all receive such votes by round $3\Delta t + \Delta$, at which point they are merged into their views. Therefore, all of their views contain $|H_t| \geq \frac{2}{3}n$ votes for $B$ from slot $t$, and $B$ is fast confirmed.
\end{proof}

\end{document}